\documentclass[12pt]{article}
\usepackage{amsmath,amssymb,epsfig}
\def\vep{\varepsilon}

\def\pipi{$\pi$-$\pi\;$}
\def\pieta{$\pi^0$-$\eta,\eta'\;$}

\def\epsrat{$\epsilon^\prime/\epsilon\;$}
\def\reeps{$\,{\rm Re}\,(\epsilon^\prime/\epsilon)\;$}
\def\optwo{${\cal O}(p^2)\;$}
\def\opfour{${\cal O}(p^4)\;$}
\def\ra{\rightarrow}

\def\be{\begin{equation}}
\def\ee{\end{equation}}
\def\bea{\begin{eqnarray}}
\def\eea{\end{eqnarray}}
\def\non{\nonumber}
\def\bb{\bibitem}

\begin{document}
\baselineskip=17pt
\parskip=5pt
  
\begin{titlepage} \footskip=5in     

\title{ 
\begin{flushright} \normalsize 
FNAL-Pub-00/136-T \\
UK/TP 00-04  \\ 
hep-ph/0006240 \\
June 2000
\vspace{5em} \\  
\end{flushright}
\large\bf The Impact of $|\Delta I|=5/2$ Transitions in 
${\bf K\ra\pi\pi}$ 
Decays}  

\author{\normalsize\bf 
S.~Gardner$^{(a)}$\thanks{E-mail: gardner@pa.uky.edu} \  and  
G.~Valencia$^{(b)}$\thanks{E-mail: valencia@iastate.edu}  \\
\normalsize\it $(a)$ Department of Physics and Astronomy, 
University of Kentucky, Lexington, KY 40506-0055\\
\normalsize\it $(b)$ Fermi National Accelerator Laboratory,
Batavia, IL 60510 \\
\normalsize\it $(b)$
Department of Physics and Astronomy, 
Iowa State University, Ames, IA 50011\footnote{Permanent Address.} \\ 
}

\date{}
\maketitle 

\begin{abstract}
We consider the impact of isospin violation on the analysis 
of $K\ra\pi\pi$ decays. We scrutinize, in particular, 
the phenomenological role played by 
the additional weak amplitude, of $|\Delta I|=5/2$ 
in character, incurred by the presence of
isospin violation. 
We show that Watson's theorem is appropriate in
${\cal O}(m_d-m_u)$, so that  
the inferred $\pi-\pi$ phase
shift at $\sqrt{s}=m_{K}$ determines the strong phase difference
between the $I=0$ and $I=2$ amplitudes in
$K\ra\pi\pi$ decay. We find the magnitude of the 
$|\Delta I|=5/2$ amplitude thus implied by the 
empirical branching ratios to be larger than expected from
estimates of isospin-violating strong and electromagnetic
effects. We effect a 
new determination of the octet and 27-plet 
coupling constants with strong-interaction isospin
violation and with electromagnetic
effects, as computed by Cirigliano, Donoghue, and Golowich,
and find that we are unable to resolve the difficulty. 
Exploring the role of $|\Delta I|=5/2$ transitions
in the CP-violating observable \epsrat, we
determine that the presence of a $|\Delta I|=5/2$ 
amplitude impacts the empirical determination of $\omega$, 
the ratio of the real parts of the $|\Delta I|=3/2$ to $|\Delta I|=1/2$ 
amplitudes, and that it generates a decrease 
in the estimation of \epsrat. 
\end{abstract}

\end{titlepage}

\section{Introduction}
\label{intro}

In the limit of 
isospin symmetry, the decay of a kaon, with isospin
$I_i=1/2$, into 
two pions, with isospin $I_f=0$ or $I_f=2$, 
is mediated by either $|\Delta I|=1/2$ or $|\Delta I|=3/2$
weak transitions. The analysis of $K\ra\pi\pi$ branching
ratios in this limit indicates that the 
$|\Delta I|=1/2$ amplitude exceeds the  $|\Delta I|=3/2$
amplitude by a factor of roughly twenty. A detailed understanding of
this large enhancement, termed the 
``$|\Delta I|=1/2$ rule,'' 
has proven elusive, although recently the subject has
received much attention~\cite{bijnens}. 
However, another, potentially related, puzzle remains. 
Unitarity and CPT invariance, in concert with isospin
symmetry, predicts that the strong 
phase difference between the $I_f=2$ and $I_f=0$ amplitudes
in $K\ra\pi\pi$ decay should equal that of 
the $I=2$ and $I=0$ amplitudes in $s$-wave $\pi\pi$ 
scattering. The 
analysis of the $K\ra\pi\pi$ branching ratios, using
isospin-symmetric amplitudes but physical phase space, 
indicates, however, that this is not the case.
Specifically, the strong phase difference inferred from $K\ra\pi\pi$ decays
is $\delta_0-\delta_2=56.6^\circ\pm 4.5^\circ$~\cite{meissner},
whereas that from $s$-wave, $\pi\pi$ scattering at the kaon mass is 
$\delta_0-\delta_2=45^\circ\pm 6^\circ$~\cite{meissner,ananth}. 

It is our purpose to examine how 
isospin-violating effects impact this apparent discrepancy.
The $u$ and $d$ quarks differ both in their charges and masses, so
that the symmetry of the $K\ra\pi\pi$ decay amplitudes 
under $u$ and $d$ quark exchange is merely
approximate. In specific, if we continue to use the labels
``$I_f=0$'' and ``$I_f=2$'' to denote the combinations of $K\ra\pi\pi$ 
amplitudes which correspond to $\pi\pi$ final states of definite isospin
in the isospin-perfect limit, then in this basis, the weak
transitions are of $|\Delta I|=1/2$, $3/2$, and $5/2$ in character. 
The violation of isospin symmetry thus generates an additional 
amplitude with $|\Delta I|=5/2$. Such effects
can modify the $|\Delta I|=1/2$ and $|\Delta I|=3/2$ amplitudes 
as well, though the large empirical enhancement of the 
$|\Delta I|=1/2$ amplitude relative to the $|\Delta I|=3/2$ amplitude
found in the isospin-conserving analysis 
suggests that isospin-violating contributions built on
the former are of greater phenomenological significance.
Indeed, it has long been suspected 
that isospin-breaking effects contaminate the
extracted ratio of $|\Delta I|=3/2$ to $|\Delta I|=1/2$ amplitudes
in a non-trivial way, precisely
as isospin violation in the ``large'' $|\Delta I|=1/2$ amplitude generates
a contribution of $|\Delta I|=3/2$ in character --- and as
the scale of strong interaction isospin violation, 
$(m_d - m_u)/m_s$, is crudely commensurate with 
that of the ratio determined in an isospin-perfect analysis. 
Indeed, including $m_d \ne m_u$ effects in a 
leading-order chiral analysis
makes the ``true'' ratio of $|\Delta I|=3/2$ to $|\Delta I|=1/2$ amplitudes
some 30\% smaller~\cite{holstein,cheng}. 
We extract the $|\Delta I|=1/2$, $3/2$, and $5/2$
amplitudes from the empirical $K\ra\pi\pi$ branching ratios
and then proceed to examine what solutions for the ``true''
$|\Delta I|=1/2$ and $3/2$ amplitudes may emerge. 

Interestingly, these considerations impact the Standard
Model (SM) estimate of \epsrat as well, for in standard practice
the empirical value of 
the ratio of the real parts of the 
$|\Delta I|=3/2$ to $|\Delta I|=1/2$ amplitudes
is used, in concert
with a ``short-distance'' determination of the 
amplitudes' imaginary parts, to determine \epsrat in the 
SM~\cite{burasr,buraslh}. 
Isospin violation plays an important role
in the analysis of
$\epsilon^\prime/\epsilon$, for it modifies the cancellation of 
the imaginary to real part ratios in the
$|\Delta I|=1/2$ and $|\Delta I|=3/2$ $K\ra \pi\pi$ amplitudes 
in a significant manner~\cite{bijwi,dght,buge,sggv,ecker}. 
The value of $\omega$, the ratio of the real parts of
the $|\Delta I|=3/2$ to $|\Delta I|=1/2$ amplitudes, 
used, however, emerges 
from an analysis of $K\ra \pi\pi$ branching ratios~\cite{devdic,kmwplb},
under the assumption that isospin symmetry is perfect. 
Thus we also explore the connection between 
isospin violation in \reeps and isospin violation 
in the $K\ra\pi\pi$ branching ratios. We determine
that the standard practice suffices to leading order 
in isospin violation if 
$|\Delta I|=5/2$ transitions can be neglected. 
The $|\Delta I|=5/2$ transitions enter differently 
in charged kaon and neutral kaon decays, and 
as the value of $\omega$ incorporated is derived, in part,
from the $K^+\ra\pi^+\pi^0$ branching ratio, the value of
$\omega$ must 
be adjusted for $|\Delta I|=5/2$ effects in order to estimate
\epsrat. This decreases the value of \epsrat and adds to its
uncertainty as well. 

We begin by considering the constraints that unitarity
and time-reversal invariance place on the parametrization
of the $K\ra\pi\pi$ amplitudes in the presence of
strong-interaction isospin violation. We consider exclusively
$m_d\ne m_u$ effects as 
electromagnetic effects are considered in Ref.~\cite{cirigliano}. 
With an appropriate parametrization in place, 
we consider the phenomenological analysis
of the $K\ra\pi\pi$ branching ratios, extracting the
amplitudes associated with the possible weak transitions 
and comparing these results with
a chiral analysis. 
We then turn to \epsrat and consider how isospin-violating
effects in the branching ratios are related to those
in \epsrat. 

\section{Unitarity Constraints}
\label{watthm}

We seek to determine what constraints may be brought to bear
on the parametrization of the $K\ra\pi\pi$ amplitudes
in the presence of isospin violation. To this end enters
Watson's theorem. We note that in 
the isospin-perfect limit, 
unitarity, and CPT invariance yields~\cite{dertasi}
\begin{eqnarray}
\langle (\pi\pi)_I | {\cal H}_W | K^0 \rangle &=& i A_I \exp(i\delta_I) \non\\
\langle (\pi\pi)_I | {\cal H}_W | \overline{K^0} \rangle &=& -i A_I^\ast \exp(i\delta_I)\,,
\label{watparam}
\end{eqnarray}
where 
${\cal H}_W$ is the effective weak Hamiltonian for kaon decays. 
The amplitude $A_I$ is such that 
$A_I = |A_I| \exp{(i\xi_I)}$, where $\xi_I$ is
the weak phase associated with the decay to the final state of 
isospin $I$, and $\delta_I$ is the phase associated with 
$s$-wave \pipi scattering of isospin $I$. 

In the limit of isospin symmetry, Bose statistics requires that
two $s$-wave pions have either $I=0$ or $I=2$. 
To relate the isospin states to 
the physical states, 
we use the isospin decomposition~\cite{foot}
\begin{eqnarray}
| \pi^+\pi^-\rangle &\propto& 
|(\pi\pi)_0 \rangle 
+ \frac{1}{\sqrt{2}}  |(\pi\pi)_2 \rangle 
\non \\
| \pi^0\pi^0\rangle 
&\propto& 
|(\pi\pi)_0 \rangle 
- {\sqrt{2}}  |(\pi\pi)_2 \rangle 
\;.
\label{isodecomp}
\end{eqnarray}
Using Watson's theorem, Eq.~(\ref{watparam}), and 
including isospin-violating effects, 
we have the parametrization
\begin{eqnarray}
A_{K^0 \ra \pi^+\pi^-}\equiv
\langle \pi^+ \pi^- | {\cal H}_W | K^0\rangle
&=& 
 i(A_0 e^{i\delta_0}
+ \frac{1}{\sqrt{2}}  A_2 e^{i\delta_2} + A_{\rm IB}^{+-} e^{i\delta_{+-}})
\non \\
A_{K^0 \ra \pi^0\pi^0}\equiv
\langle \pi^0 \pi^0 | {\cal H}_W | K^0\rangle
&=& 
i( A_0 e^{i\delta_0}
- {\sqrt{2}}  A_2 e^{i\delta_2} + A_{\rm IB}^{00} e^{i\delta_{00}}) \;, 
\label{trueparam} \\
A_{K^+ \ra \pi^+\pi^0}\equiv
\langle \pi^+ \pi^0 | {\cal H}_W | K^+\rangle
&=&
i(
\frac{3}{2} A_2 e^{i\delta_2} + A_{\rm IB}^{+0} e^{i\delta_{+0}}) \;,
\non
\end{eqnarray}
where the isospin-violating contributions are denoted by the subscript
``${\rm IB}$'' and include a weak phase, e.g., 
 $A_{\rm IB}^{00}= |A_{\rm IB}^{00}| e^{i\xi_{00}}$. The
strong phases $\delta_{00}$, $\delta_{+-}$, and
$\delta_{+0}$ are, as yet, 
idiosyncratic to $K\ra \pi\pi$ decay.
As $A_0$ and $A_2$ are reflective of the amplitudes
in the isospin-perfect limit, they are generated by 
$|\Delta I|=1/2$ and $|\Delta I|=3/2$ weak transitions, respectively. 

We wish to examine what further constraints may be placed
on Eq.~(\ref{trueparam}). It follows from unitarity that 
a transition matrix $T$ 
satisfies the relation
\begin{equation}
   T^\dagger T = i (T^\dagger -  T) \;,
\end{equation} 
where the $S$ matrix can be written as $S = 1 + i T$ and unitarity is the
condition $S^\dagger S =1$. We consider
$K \ra \pi \pi$ decays, so that the final-state phases of interest
are generated through $\pi$-$\pi$ scattering. In the presence of
isospin violation, the isospin-perfect basis of Eq.~(\ref{isodecomp})
continues to prove convenient, 
as the possibility of 
$\pi^+\pi^- \leftrightarrow \pi^0 \pi^0$ through strong 
rescattering makes the ``physical'' basis awkward. 
The label ``$I$,'' however, need only correspond to  the isospin of
the final-state pions in the isospin-perfect limit.
We begin by considering $K^0\ra (\pi\pi)_I$ decays and 
find, upon insertion of 
{\it all} possible intermediate states $F$:
\begin{equation}
\sum_F \langle 
(\pi\pi)_I |   T^\dagger | F \rangle \langle F | T| K^0\rangle 
= i (\langle (\pi\pi)_I | 
T^\dagger| K^0\rangle  -  \langle (\pi\pi)_I | T | K^0\rangle) \;.
\label{watson}
\end{equation}
Note that $F$ denotes the set of states physically accessible in $K$
decay and thus includes the 
$(\pi\pi)_I$ states defined in Eq.~(\ref{isodecomp}), as well as
$\pi^+\pi^-\gamma$,
$\gamma\gamma$, and $3\pi$ states. 
In the isospin-perfect limit, only the $F=(\pi\pi)_I$ 
term in the sum contributes.  The inclusion of
electromagnetic effects, however, complicates matters, as 
additional states 
may contribute to the sum in Eq.~(\ref{watson}). 
The most significant of the modes with photons or leptons in the
final state is $K_S^0\ra\pi^+\pi^-\gamma$;
let us continue to neglect such electromagnetic 
isospin-violating effects and investigate 
the effects of strong-interaction isospin violation. 
We also neglect the 
3$\pi$ intermediate state appearing in Eq.~(\ref{watson}) because the 
$\langle (\pi\pi)_I | T | 3 \pi\rangle$ 
transition amplitude with $J=0$ violates not only 
G-parity but P as well. Note that the spatial component of the
$J=0$ 3$\pi$ state 
is even under $P$, so that the $J=0$ 3$\pi$ state is of
odd parity~\cite{oddpar}.  
We work to leading order in the weak interaction, so that
$\langle 2\pi | T | 3\pi \rangle$ is mediated by strong rescattering
and thus vanishes for $J=0$ states, as the strong interaction
conserves parity.  
At the energies appropriate to kaon decay, 
the strong scattering in the $(\pi\pi)_I$ final state 
is described by a pure 
phase, as the empirical inelasticity parameters are unity~\cite{roy}, 
so that in the isospin-perfect limit we can write
\begin{equation}
{\bf S} = 
\begin{pmatrix}{e^{2i\delta_0}} & 0 \\ 0  & {e^{2i\delta_2}}\end{pmatrix}
\;.
\label{smatrixiso}
\end{equation}
Thus if isospin is a perfect symmetry, 
only $F=(\pi\pi)_I$ contributes to the sum and one recovers
the usual parametrization 
\begin{eqnarray}
\langle (\pi\pi)_I |  T | K^0 \rangle &=& i A_I \exp(i\delta_I) \non \\
\langle (\pi\pi)_I |  T | \overline{K^0} \rangle 
&=& -i A_I^\ast \exp(i\delta_I)\;,
\label{watparamT}
\end{eqnarray}
noting by CPT symmetry that $\langle (\pi\pi)_I |  T^\dagger | K^0 \rangle = 
(\langle (\pi\pi)_I |  T | \overline{K^0} \rangle)^\ast$. 

We now turn to the consideration of isospin-violating effects. The 
$S$-matrix appropriate to the $\pi\pi$ final states with zero net charge
is characterized, in general, by eight real parameters.
Unitarity, however, yields three distinct constraints, and 
time-reversal invariance yields two more, so that the $S$-matrix
can contain at most three real parameters. We have seen from the
explicit form of $S$-matrix in the isospin-perfect limit that
it is characterized by precisely two parameters, $\delta_0$
and $\delta_2$ --- and thus the third parameter permitted by
unitarity and time-reversal invariance must be at least 
of ${\cal O}(m_d-m_u)$, or of ${\cal O}(\alpha)$. As electromagnetic
effects in the $K\ra \pi\pi$ phases are studied 
in Ref.~{\cite{cirigliano}}, we focus on $m_d\ne m_u$ effects. 

We parametrize the $S$-matrix 
in the presence of isospin violation
as~\cite{preston} 
\begin{equation}
{\bf S} = 
\begin{pmatrix}{e^{i{\bar \delta}_0}} & 0 \\ 0  & {e^{i{\bar \delta}_2}}
\end{pmatrix}
\begin{pmatrix} \cos 2\kappa & i\sin 2\kappa \\ i\sin 2\kappa & \cos 2\kappa 
\end{pmatrix}
\begin{pmatrix}{e^{i{\bar \delta}_0}} & 0 \\ 0  & {e^{i{\bar \delta}_2}}
\end{pmatrix}
\label{smatrix}
\end{equation}
where the third S-matrix parameter is denoted by $\kappa$. 
Note that if $\kappa=0$ then 
$\bar{\delta}_I=\delta_I$, where $\delta_I$ 
denote the strong phases of the isospin-perfect limit. 
In the presence of isospin violation we continue 
to use  Eq.~(\ref{isodecomp}) 
to define the $|(\pi\pi)_I\rangle$ states 
used in Eq.~(\ref{smatrix}). 
The parameter $\kappa$ is 
sensitive to $m_d\ne m_u$ effects in the strong chiral
Lagrangian, as well as to electromagnetic effects. 
Explicit calculation shows that all 
strong-interaction isospin-violating effects in $\pi\pi$ scattering
are at least of ${\cal O}((m_d - m_u)^2)$ in \opfour in the chiral
expansion~\cite{glann}. This result persists to all orders
in chiral perturbation theory; let us turn to an explicit 
demonstration of this point. 

Isospin violation in the S-matrix element for
2-to-2 $\pi\pi$ scattering can occur in either
the truncated, connected Green function itself or in the external
$\pi$ legs. The latter source of isospin violation 
emerges as in ${\cal O}(m_d-m_u)$
the $\pi^0$ and $\eta$ fields mix. Diagonalizing the neutral,
non-strange meson states of the strong chiral Lagrangian yields,
in ${\cal O}(p^2)$, e.g., yields the ``physical''
$\pi^0$ state in terms of the pseudoscalar octet fields $\pi^0$ and 
$\eta$~\cite{gl}:
\begin{equation}
\left(\pi^0\right)_{\rm phys}
= \pi^0 + {\frac{\sqrt{3}}{4}}\left({\frac{m_d-m_u}{m_s -\hat{m}}}\right)\eta 
+ {\cal O}((m_d - m_u)^2)\;,
\label{piphys}
\end{equation}
where $\hat{m}=(m_d+m_u)/2$. An analogous formula exists in \opfour~\cite{gl}.
Thus isospin violation in an external $\pi$ leg is realized
as an $\eta$ admixture in the physical $\pi^0$ state. 
In the pseudoscalar octet, or ``isospin-perfect,'' basis we have
adopted thus far, an ${\cal O}(m_d-m_u)$ interaction converts
the isovector $\pi^0$ into a isoscalar $\eta$. Thus 
in ${\cal O}(m_d-m_u)$ 
the truncated, connected Green function arising from isospin
violation in an external $\pi$ leg 
contains one $\eta$ and three $\pi$ fields.
Note that the decay $\eta\ra\pi\pi\pi$ is 
forbidden by Bose symmetry in the isospin-symmetric
limit, $m_d=m_u$, so that the truncated, connected Green function 
of interest must be at least of ${\cal O}(m_d-m_u)$. Including the 
$(m_d-m_u)$ ``penalty'' required to convert the $\eta$ to a physical
$\pi^0$, one finds that isospin-violating effects arising from
the external legs start in ${\cal O}((m_d-m_u)^2)$. One
can also show that the $m_d\ne m_u$ effects 
in the truncated, connected Green function associated with
the 2-to-2 scattering of isovector pions also start in
${\cal O}((m_d-m_u)^2)$. 
Following the ``spurion'' formulation~\cite{tdlee}, 
a transition matrix element with SU(2) violation must
have the same properties as a SU(2)-conserving transition
matrix element containing a spurion,
a fictitous particle which carries, in this case,
the quantum numbers of the $\pi^0$ and 
a factor of $(m_d-m_u)$. Thus the spurion and the $\pi$ are both of 
negative G-parity, so that a transition of form 
\begin{equation}
\hbox{(even number of pions)} \Longleftrightarrow 
\hbox{(even number of pions + 1 spurion)}
\end{equation}
is forbidden by G-parity and does not occur~\cite{clebsch}. 
Note, however, that a transition of form 
\begin{equation}
\hbox{(even number of pions)} \Longleftrightarrow 
\hbox{(even number of pions + 2 spurions)}
\end{equation}
is permitted by G-parity, so that all isospin-violating
effects in $\pi$-$\pi$ scattering are of ${\cal O}((m_d - m_u)^2)$.
Analyzing Eq.~(\ref{smatrix}), this result implies
that 
\begin{equation}
{\bar \delta}_I - \delta_I \sim {\cal O}((m_d - m_u)^2) \quad ; \quad
\kappa \sim {\cal O}((m_d - m_u)^2) \;.
\label{plimit}
\end{equation}
so that $\kappa=0$ in ${\cal O}(m_d-m_u)$. 

Using Eq.~(\ref{smatrix}) to incorporate isospin violation 
in $K\ra\pi\pi$ decays, we find that 
Eq.~(\ref{watson}) thus becomes
\begin{equation}
\begin{pmatrix}
{1 - e^{-2i{\bar \delta}_0} \cos 2\kappa } &
{-i e^{-i({\bar \delta}_0+{\bar \delta}_2)} \sin 2\kappa }
\\ 
{-i e^{-i({\bar \delta}_0+{\bar \delta}_2)} \sin 2\kappa  
} & 
{1 - e^{-2i{\bar \delta}_2} \cos 2\kappa }
\end{pmatrix}
\begin{pmatrix} 
{\langle (\pi\pi)_0 | T | K^0 \rangle}  \\ 
{\langle (\pi\pi)_2 | T | K^0 \rangle}
\end{pmatrix}
=
\begin{pmatrix} 
{\langle (\pi\pi)_0 | T | K^0 \rangle - 
\langle (\pi\pi)_0 | T^\dagger | K^0 \rangle}
\\ 
{\langle (\pi\pi)_2 | T | K^0 \rangle - 
\langle (\pi\pi)_2 | T^\dagger | K^0 \rangle}
\end{pmatrix}
\label{coupled1}
\end{equation}
Following the parametrization of Eq.~(\ref{watparamT}), we have
in the presence of isospin violation
\begin{eqnarray}
\langle (\pi\pi)_I | T | K^0 \rangle &=& 
i A_I \exp(i\tilde\delta_I) \non\\
\langle (\pi\pi)_I | T | \overline{K^0} \rangle &=& 
-i A_I^\ast \exp(i\tilde\delta_I)\,,
\label{watparamIB}
\end{eqnarray}
where $\tilde\delta_I$, the strong phase of the $K\ra\pi\pi$
decay amplitude, is related to the strong phase of $\pi\pi$
scattering, given in Eq.~(\ref{smatrix}), 
as per Eq.~(\ref{coupled1}). We thus have
\begin{equation}
\begin{pmatrix}
{1 - e^{-2i{\bar \delta}_0} \cos 2\kappa } &
{-i e^{-i({\bar \delta}_0+{\bar \delta}_2)} \sin 2\kappa }
\\ 
{-i e^{-i({\bar \delta}_0+{\bar \delta}_2)} \sin 2\kappa  
} & 
{1 - e^{-2i{\bar \delta}_2} \cos 2\kappa }
\end{pmatrix}
\begin{pmatrix} 
A_0 e^{i\tilde\delta_0}
\\ 
A_2 e^{i\tilde\delta_2}
\end{pmatrix}
\quad=\quad
2i
\begin{pmatrix} 
A_0 \sin\tilde\delta_0
\\ 
A_2 \sin\tilde\delta_2
\end{pmatrix}
\label{coupled2}
\end{equation}
Note that if the channel-coupling parameter $\kappa$ were zero, 
then ${\tilde\delta_I} = {\bar \delta_I}=\delta_I$, and the
strong-phase in the $K\ra\pi\pi$ decay amplitude would be 
that of $\pi\pi$ scattering, analyzed in the isospin-perfect limit.
Defining 
\begin{equation}
\Delta_I \equiv {\bar \delta}_I - \tilde\delta_I\;,
\end{equation}
so that $\Delta_I=0$ were $\kappa=0$,
and rearranging the upper component of Eq.~(\ref{coupled2}), 
we find
\begin{equation}
e^{-2 i \Delta_0} \cos (2 \kappa) - 1 
= 
-i \frac{A_2}{A_0} e^{-i(\Delta_0+\Delta_2)} \sin (2\kappa) \;.
\label{coupled3}
\end{equation}
Using the lower component of Eq.~(\ref{coupled2})
yields  Eq.~(\ref{coupled3}) with the isospin subscripts
switched, $0 \leftrightarrow 2$. As $\kappa \ra 0$,
$\Delta_I\ra 0$ as well, and we find
\begin{equation}
\Delta_0 = \frac{A_2}{A_0} \kappa + {\cal O}(\kappa^2) 
\quad ; \quad
\Delta_2 = \frac{A_0}{A_2} \kappa + {\cal O}(\kappa^2) \;,
\label{dlimit}
\end{equation}
implying $\Delta_2 \gg \Delta_0$ and 
$\Delta_0 \Delta_2 \sim \kappa^2$. 
Eliminating $A_2/A_0$ from Eq.~(\ref{coupled3}) 
and its $0 \leftrightarrow 2 $ counterpart 
yields a relation purely in terms of $\Delta_I$ and $\kappa$: 
\begin{equation}
\cos (2\kappa)\cos (\Delta_0 - \Delta_2) 
= \cos (\Delta_0 + \Delta_2) \;.
\end{equation}
Alternatively, one can eliminate $\kappa$ to find
\begin{equation}
A_2^2 \sin(2\Delta_2) = A_0^2 \sin(2\Delta_0) \;.
\end{equation}
With Eqs.~(\ref{plimit}) and (\ref{dlimit}) we have that 
${\tilde \delta}_I  - \delta_I$ is no larger than
\begin{equation}
{\tilde \delta}_I - \delta_I \sim {\cal O}((m_d - m_u)^2) \;.\;
\end{equation}
Thus in ${\cal O}(m_d - m_u)$ the channel-coupling parameter 
$\kappa=0$ and ${\tilde\delta}_I=\delta_I$, so that
the parametrization of Eq.~(\ref{watparamT})
is appropriate in the presence of strong-interaction isospin
violation as well. However, if electromagnetic effects were
included, one would expect $\kappa \sim {\cal O}(\alpha)$,
and with $A_2/A_0 \sim 1/20$, one finds 
$|\Delta_2| \sim |{\tilde \delta}_2 - \delta_2|
\sim {\cal O}(10^\circ)$~\cite{neglectem}, commensurate with 
the explicit estimate of $4.5^\circ$ in ${\cal O}(e^2 p^0)$ in 
Ref.~\cite{cirigliano}.

We consider how our results generalize to the case of
$K^+\ra\pi^+\pi^0$ decays as well, for these decays are needed
to isolate the $|\Delta I|=5/2$ amplitude.
In the case of charged $K\ra \pi\pi$ decays, 
Eq.~(\ref{watson}) becomes
\begin{equation}
\sum_F \langle (\pi\pi)_{I^+}| T^\dagger | F \rangle \langle F | T| K^+\rangle 
= i (\langle (\pi\pi)_{I^+} | T^\dagger| K^+\rangle  
-  \langle (\pi\pi)_{I^+} | T | K^+\rangle) \;,
\label{watsonp}
\end{equation}
where we now explicitly denote the isospin $I$, $I_3=1$ final
state by ``$(\pi\pi)_{I^+}$''. Charge is conserved so that Eq.~(\ref{watsonp})
is diagonal in $I_3$. Neglecting the $3\pi$ and electromagnetic
intermediate states, we thus have
\begin{equation}
\langle (\pi\pi)_{2^+} |   T^\dagger | (\pi\pi)_{2^+} \rangle 
\langle (\pi\pi)_{2^+} |   T | K^+ \rangle 
= i (\langle (\pi\pi)_{2^+} | T^\dagger| K^+\rangle  - 
 \langle (\pi\pi)_{2^+} | T | K^+\rangle) 
\;.
\label{watson2p}
\end{equation}
By crossing symmetry, 
our prior analysis of isospin violation
in $\pi\pi$ scattering is germane to this case as well, so that 
we conclude that strong-interaction
isospin-violating effects in 
$\langle (\pi\pi)_{2^+}| T^\dagger | (\pi\pi)_{2^+} \rangle$
are of ${\cal O}((m_d-m_u)^2)$. Thus we write 
$\langle (\pi\pi)_{2^+} | T^\dagger | (\pi\pi)_{2^+} \rangle = 
-i(1 - e^{-2i\delta_2})$, or
finally 
\begin{equation}
\langle (\pi\pi)_{2^+} | T | K^+ \rangle = i A_{2^+} e^{i\delta_2} \;,
\label{chargedk}
\end{equation}
so that, with the neglect of electromagnetic effects,
the strong phase in this channel is related to that 
of the $I=2$ amplitude comprised of 
charge-neutral final states.  It is worth noting
that the phase of Eq.~(\ref{chargedk}) is
evaluated at $\sqrt{s}=m_{K^+}$, whereas the 
phases of $K^0\ra\pi\pi$ decay is evaluated
at $\sqrt{s}=m_{K^0}$. However, this small difference 
is without practical consequence, for the phase of Eq.~(\ref{chargedk}) 
does not appear in the $K^+\ra \pi\pi$ branching ratio.

We have thus demonstrated in ${\cal O}(m_d-m_u)$ that
the strong phases of the $K\ra\pi\pi$ amplitudes are those
of $\pi\pi$ scattering in the isospin-perfect
limit. Generally, $m_d\ne m_u$ effects permit 
amplitudes of $|\Delta I|=1/2$, $3/2$, and $5/2$ in character, so 
that the parametrization of Eq.~(\ref{trueparam}) can be rewritten as
\begin{eqnarray}
A_{K^0 \ra \pi^+\pi^-}
&=& 
i((A_{0} + \delta A_{1/2})e^{i\delta_0}
+ \frac{1}{\sqrt{2}}  (A_{2} + \delta A_{3/2} + \delta A_{5/2}) 
e^{i\delta_2} ) 
\non \\
A_{K^0 \ra \pi^0\pi^0}
&=& 
i( (A_{0} + \delta A_{1/2})e^{i\delta_0}
- {\sqrt{2}}  (A_{2} + \delta A_{3/2} + \delta A_{5/2}) e^{i\delta_2} )
\;
\label{newparam} \\
A_{K^+ \ra \pi^+\pi^0}
&=& 
i(\frac{3}{2}(A_2 + \delta A_{3/2}) - \delta A_{5/2}) e^{i\delta_2} 
\;,
\non
\end{eqnarray}
in ${\cal O}(m_d-m_u)$, where
$\delta A_{|\Delta I|}$ denotes the amplitude contributions induced
exclusively by isospin violation. Note that the parametrization
of the charge-conjugate decays follows from Eq.~(\ref{watparamIB}).
The $\delta A_{1/2}$ and 
$\delta A_{3/2}$ contributions are each generated by both 
$|\Delta I|=1/2$ and $|\Delta I|=3/2$
weak transitions. The presence of a $\delta A_{5/2}$ contribution
--- the ``new'' amplitude --- 
is signalled by the inequality
$(A_{K^0 \ra \pi^+\pi^-} - A_{K^0 \ra \pi^0\pi^0})/\sqrt{2} -
A_{K^+ \ra \pi^+\pi^0} \ne 0$~\cite{sgpieta}. 

\section{Phenomenology of $K\ra\pi\pi$ Decays} 
\label{pheno}

We now wish to determine the relative magnitude of the
various amplitudes in Eq.~(\ref{newparam}) 
predicated by the measured $K\ra\pi\pi$ 
branching ratios and by the inferred \pipi phase shifts. To this end,
we consider the following ratios of reduced transition rates: 
\begin{equation}
R_1=
\frac{\gamma(K_S^0\ra\pi^+\pi^-)}{\gamma(K_S^0\ra\pi^0\pi^0)}\; 
\label{rat1}
\end{equation}
and
\begin{equation}
R_2=
\frac{2\gamma(K^+\ra\pi^+\pi^0)}
{\gamma(K_S^0\ra\pi^+\pi^-)+\gamma(K_S^0\ra\pi^0\pi^0)}\;,
\label{rat2}
\end{equation}
where $\gamma(K\ra\pi_1\pi_2)$, the reduced transition rate, 
is related to the partial width $\Gamma(K\ra\pi_1\pi_2)$ via
\begin{equation}
\Gamma(K\ra\pi_1\pi_2)= \frac{\sqrt{(m_K^2 - (m_{\pi_1} + m_{\pi_2})^2)
(m_K^2 - (m_{\pi_1} - m_{\pi_2})^2)}}{16\pi m_K^3}
\gamma(K\ra\pi_1\pi_2) \;.
\end{equation}
We use the physical $\pi$ and $K$ masses in extracting
$\gamma(K\ra\pi\pi)$, and neglect any final-state
Coulomb corrections as they are electromagnetic effects.
The reduced transition rates are simply related to the absolute
squares of the amplitudes we have considered previously, so that
\begin{eqnarray}
&&R_1=
\frac{2 |A_{K_S^0\ra\pi^+\pi^-}|^2}{|A_{K_S^0\ra\pi^0\pi^0}|^2} \non \\
&&R_2=\frac{2|A_{K^+\ra\pi^+\pi^0}|^2}
{2 |A_{K_S^0\ra\pi^+\pi^-}|^2 + |A_{K_S^0\ra\pi^0\pi^0}|^2}\;.
\end{eqnarray}
Using the parametrization
of Eq.~(\ref{newparam}), noting
$K_S=(K^0 - \overline{K^0})/\sqrt{2}$ with $CP(K^0)=-\overline{K^0}$,
while ignoring CP violation and weak phases, 
yields
\begin{eqnarray}
2\sqrt{\frac{R_2}{3}}&=& \pm ( x -\frac{2}{3} y)  \;;
\label{R2xy}\\
\frac{R_1}{2}&=&
\frac{1 + \sqrt{2}(x+y) \cos(\delta_2-\delta_0) + (x+y)^2/2}
{1 - 2\sqrt{2}(x+y) \cos(\delta_2-\delta_0) + 2(x+y)^2} 
\non \\
&=& 1 + 3\sqrt{2}(x+y)\cos(\delta_2-\delta_0) + 
(12\cos^2(\delta_2-\delta_0) - 3/2)x^2 + {\cal O}(xy,x^3,y^2) 
\label{R1xy} \;, 
\end{eqnarray}
where, working consistently to leading order in isospin violation,
we have
\begin{eqnarray}
x&\equiv& \frac{A_{2} + \delta A_{3/2}}{A_{0} + \delta A_{1/2}} \approx
\frac{A_{2}}{A_{0}} + \frac{\delta A_{3/2}}{A_{0}} - 
\frac{A_{2}}{A_{0}}\frac{\delta A_{1/2}}{A_{0}} \;, \non\\
y&\equiv& \frac{\delta A_{5/2}}{A_{0} + \delta A_{1/2}} \approx
\frac{\delta A_{5/2}}{A_{0}} 
\label{xydef} \;. 
\end{eqnarray}
The ratio $x$ is $A_2/A_0$ in the isospin-perfect limit, 
whereas the ratio $y$ is non-zero only in the presence of
isospin violation. 
We anticipate that a $\delta A_{5/2}$ contribution is 
generated either by strong-interaction isospin
violation in concert with a $|\Delta I|=3/2$ weak transition, 
or by electromagnetic effects in concert with a 
$|\Delta I|=1/2$ weak transition.  
We thus expect the hierarchy $x \gg x^2, y \gg x^3,xy,y^2$, which is
reflected in the terms retained in Eq.~(\ref{R1xy}). 
Note that it is appropriate to continue to work to leading
order in isospin violation after the inclusion of the 
$|\Delta I|=5/2$ contributions, as crudely $|A_2/A_0| \sim 5 \%$ 
--- this follows from Eq.~(\ref{R2xy}) if $y=0$ ---
whereas isospin violation is a $\sim 1\%$ effect. 

\begin{figure}[b!] 
\vspace{100pt}
\centerline{\epsfig{file=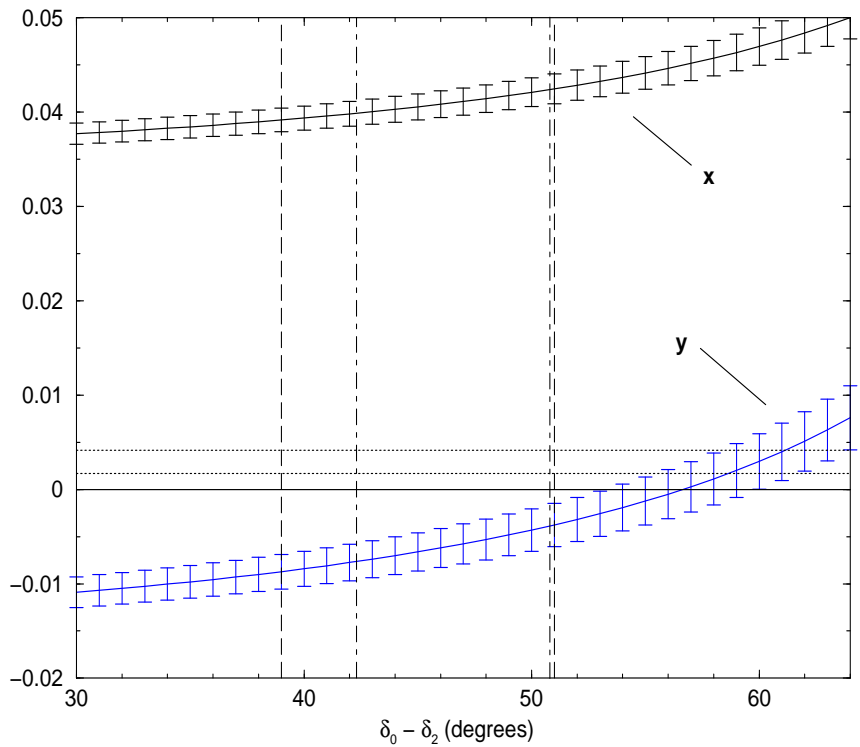,height=3.75in,width=2.0in,angle=-90}}
\vspace{100pt}
\caption{
The values of $x$ and $y$ resulting from 
Eqs.~(\protect{\ref{R2xy}},\protect{\ref{R1xy}}) as a function
of $\delta_0-\delta_2$. In the isospin-perfect limit 
$x=A_2/A_0$ and $y=0$. 
The vertical dashed and dot-dashed lines enclose the results 
\protect{$\delta_0-\delta_2=45^\circ \pm 6^\circ$~\cite{meissner}}
and
\protect{$\delta_0-\delta_2=45.2^\circ \pm 1.3^\circ \pm
\stackrel{4.5^\circ}{{\;\!}_{1.6^\circ}}$~\cite{ananth}}, respectively, 
at 68\% C.L. The two sets of vertical lines overlap
at $51^\circ$ --- the dot-dashed line has been slightly
off-set for presentation. 
The horizontal dashed line encloses the electromagnetic
contribution to $y$ as per the ``dispersive 
matching'' calculation of Table I of 
\protect{Ref.~\cite{cirigliano}} at 68\% C.L. 
}
\label{fig1}
\end{figure}
Let now proceed to determine $x$ and $y$. 
We determine $R_1$ and $R_2$ using the ``our fit'' branching
ratios and ancillary empirical data in Ref.~{\cite{pdg98}}
and plot the 
$x$ and $y$  resulting from Eqs.~(\ref{R2xy},\ref{R1xy})  
as a function of 
$\delta_0-\delta_2$ in Fig.~1. Note that 
$\cos(\delta_2 - \delta_0)>0$ and $R_1/2>1$, so that Eq.~(\ref{R1xy})
implies that $x+y>0$. As we assume $x\gg y$, then $x>0$
as well, and we choose the $+$ sign in Eq.~(\ref{R2xy})
in what follows~\cite{altsgn}. 
Moreover, 
we pick the root of the quadratic equation consistent with 
$A_0> A_2$. We affect these choices in order to recover
the qualitative features of the analysis performed in 
the $m_d\ra m_u$ limit. The errors in $x$ and $y$
arise from the empirical errors, assuming all the errors
are uncorrelated. The vertical dashed lines enclose
the phase shift difference
$\delta_0-\delta_2=45^\circ \pm 6^\circ$~\cite{meissner},
whereas the vertical dot-dashed lines enclose 
$\delta_0-\delta_2=45.2^\circ \pm 1.3^\circ \pm
\stackrel{4.5^\circ}{{\;\!}_{1.6^\circ}}$~\cite{ananth} at 68\% C.L. 
We omit explicit use of this latter value in what
follows as it is comparable to the result
of Ref.~\cite{meissner}. 
Table~\ref{tablexy} 
shows the specific values of $x$ and $y$, with their associated
errors, which emerge from combining the empirical values of
$R_1$ and $R_2$ with the values of $\delta_0 -\delta_2$ 
from various sources. 
Note that we use the $\delta_0 - \delta_2$ phase shift 
as extracted in the isospin-symmetric limit, as 
strong-interaction isospin-violating
effects enter merely in ${\cal O}((m_d - m_u)^2)$ and 
as the electromagnetically generated $K\ra \pi\pi$
phase shifts appear to be small~\cite{cirigliano}. 
For estimates of electromagnetic effects in 
$\pi-\pi$ scattering, see Ref.~\cite{emcalc}. 

Proceeding with the numerical analysis, 
we find a substantial 
value for $\delta A_{5/2}$, 
suggesting the phenomenological hierarchy 
$x \gg y \gg x^2, xy$. 
Specifically, we find
\begin{equation}
\delta A_{5/2}/(A_2 + \delta A_{3/2}) \sim 20\%\;, 
\end{equation}
rather than the
${\cal O}(1\%)$ we might have anticipated from strong-interaction
isospin violation. 
The extracted $\delta A_{5/2}$ amplitude is sensitive to the
value of $\delta_0 - \delta_2$ used; indeed, were
 $\delta_0 - \delta_2 \sim 56.6^\circ$, then 
$\delta A_{5/2} \sim 0$. Moreover, if the errors 
in $\delta_0 -\delta_2$ were consistently --- and substantially ---
underestimated, our determined $\delta A_{5/2}$ could be made 
consistent with zero. In particular, if we were to increase the error 
in $\delta_0 -\delta_2$ to realize this, we would find that
we would require, e.g., $45^\circ \pm 16^\circ$. 
Such increases would reflect a severe inflation of the stated
error bars and would seem unwarranted. It ought be realized 
that $\pi\pi$ phase shift information is largely inferred
from associated production in $\pi N$ reactions and that 
any possible theoretical systematic errors incurred 
through the choice of reaction model are not incorporated 
in the reported error estimates~\cite{sevior}. However,
information on the $I=0$ $\pi\pi$ phase shift near threshold 
is also known from 
$K\ra\pi\pi e \nu$ decay; this is consistent
with the phase shift determined  in $\pi N$ reactions, albeit
the errors are large~\cite{amoros}. 
Interestingly, the 
$e^+ e^-\ra \pi\pi$ and $\tau\ra\pi\pi\nu$ data in the
context of a Roy equation analysis of $\pi\pi$ scattering
constrain the possible $s$-wave phase shifts rather
significantly, yielding at $s=m_{K^0}^2$ that 
$\delta_0-\delta_2=45.2^\circ \pm 1.3^\circ \pm
\stackrel{4.5^\circ}{{\;\!}_{1.6^\circ}}$~\cite{ananth}.
This is commensurate with earlier determinations
of $\delta_0-\delta_2$~\cite{ochs2,gvphase,chell,devdic}, 
noting Table~\ref{tablexy}, and encourages us to consider the 
consequences of our fit. 

\begin{table}[htb]
\begin{center}
\caption{
The values of $x$ and $y$ resulting from 
Eqs.~(\protect{\ref{R2xy}},\protect{\ref{R1xy}}) 
using the phase shift differences, $\delta_0-\delta_2$, 
compiled from various sources. 
Note that in the isospin-perfect limit that 
$x=A_2/A_0$ and $y=0$. The values found for $|y|$
are roughly equal to $\alpha$, suggesting 
an electromagnetic origin for $y$. 
\smallskip
}
\begin{tabular}{cccc}
\hline
Ref. & $\delta_0-\delta_2$ (deg.) & $x$ & $y$ \\
\hline
\protect{\cite{devdic}} 
& $41.4 \pm 8.1$ & $0.0396 \pm 0.0022$ & $-0.0080 \pm 0.0033$ \\
\protect{\cite{chell}} 
& $42 \pm 4$ & $0.0398 \pm 0.0016$ & $-0.0077 \pm 0.0024$ \\
\protect{\cite{chell}} (``local fit'')
 & $42 \pm 6$ & $0.0398 \pm 0.0019$ & $-0.0077 \pm 0.0028$ \\
\protect{\cite{ochs2}}
 & $44 \pm 5$ & $0.0403 \pm 0.0019$ & $-0.0070 \pm 0.0028$ \\
\protect{\cite{meissner}}
 & $45 \pm 6$ & $0.0405 \pm 0.0021$ & $-0.0066 \pm 0.0032$ \\
\end{tabular}
\label{tablexy}
\end{center}
\end{table}

Let us first compare our results with the 
$\delta A_{5/2}$ amplitude estimated to be induced by 
electromagnetism~\cite{cirigliano}. 
Using Eq.~(48) and the ``dispersive matching'' estimate of 
Table I in Ref.~\cite{cirigliano}, we find 
$y_{\rm em}\sim 0.0029$, suggesting that the computed electromagnetic 
effects are rather smaller and are of the wrong sign~\cite{circomm}. 
Indeed, this discrepancy prompts our consideration of
strong-interaction isospin-violating effects. In particular,
were $y=0$, then Eq.~(\ref{R1xy}) would become
\begin{eqnarray}
\frac{R_1}{2}&=&
\frac{1 + \sqrt{2}x \cos(\delta_2-\delta_0) + x^2/2}
{1 - 2\sqrt{2}x \cos(\delta_2-\delta_0) + 2x^2} 
\non \\
&=& 1 + 3\sqrt{2}x\cos(\delta_2-\delta_0) + 
(12\cos^2(\delta_2-\delta_0) - 3/2)x^2 + {\cal O}(x^3) 
 \;. 
\end{eqnarray}
Using 
$\delta_0-\delta_2=45^\circ$~\cite{meissner} and 
Ref.~\cite{pdg98}  
yields $x=0.035$, whereas Eq.~(\ref{R2xy}) in this limit 
would be 
\begin{equation}
2\sqrt{\frac{R_2}{3}}= x  \;
\end{equation}
and yields $x=0.045$ --- this discrepancy is reconciled
through the value of $y$ we report in Table~\ref{tablexy}. 
The significance of $y$ could be exacerbated by the parameters
reported in Ref.~\cite{pdg98}, though
excursions of several standard deviations are required
to impact its value significantly~\cite{brvary}.

We summarize this section with the following observations.

\begin{itemize}

\item The value of $x$ is stable
with respect to the various values of $\delta_0-\delta_2$
reported in Table \ref{tablexy} --- it varies merely at the 1\% level.

\item The value of $y$ is rather more sensitive
to $\delta_0-\delta_2$. It apparently is of ${\cal O}(\alpha)$, 
rather than of ${\cal O}(\omega(m_d-m_u)/m_s)$ --- and thus is
rather larger than expected from the standpoint
of strong-interaction isospin violation. 

\end{itemize}

\section{Isospin Violation and the $|\Delta I|=1/2$ Rule}
\label{isodelta}

Our determined $x$ and $y$ may be connected to the 
amplitudes of the isospin-perfect limit, $A_{0}$ and $A_{2}$,
via a computation of the $K\ra\pi\pi$ amplitudes in 
chiral perturbation theory. The weak chiral Lagrangian
in \optwo has two non-trivial terms, which transform as 
$(8_L,1_R)$ and as $(27_L,1_R)$ 
under $SU(3)_L\times SU(3)_R$,
respectively~\cite{cronin}. We wish to determine
their relative magnitude in the context of a calculation
which is sensitive to $m_u\ne m_d$ effects, in order to 
assess the relative strength of the $(27_L,1_R)$ and
$(8_L,1_R)$ transitions, that is, the ratio $A_2/A_0$. We believe
that $m_u\ne m_d$ effects likely contribute to $x$ in a
significant manner~\cite{holstein,cheng}. Ultimately
we will also include the 
computed electromagnetic corrections of Ref.~\cite{cirigliano} as
well, in order to determine $A_2/A_0$, for the numerical value of
$y$ is crudely an ${\cal O}(\alpha)$ effect.

In ${\cal O}(p^2)$, the $(8_L,1_R)$ 
term in the weak, chiral Lagrangian generates
exclusively $|\Delta I|=1/2$ transitions, whereas the
$(27_L,1_R)$ term generates both $|\Delta I|=1/2$ and
$|\Delta I|=3/2$ transitions. 
We have~\cite{pichder91}
\begin{eqnarray}
{\cal L}_W^{(2)} &=& 
-\frac{G_F}{\sqrt{2}}
V_{ud} V_{us}^{\ast} 
\Big[
g_8 \left(L_{\mu}L^{\mu}\right)_{23}
+ g_{27}^{(1/2)} 
\left(L_{\mu 13}L_{21}^{\mu} + L_{\mu 23}(4L_{11}^{\mu} + 5L_{22}^{\mu})
\right) \non\\
&+& g_{27}^{(3/2)} 
\left(L_{\mu 13}L_{21}^{\mu} + L_{\mu 23}(L_{11}^{\mu} - L_{22}^{\mu})
\right)
\Big] + \hbox{h.c.}\;,
\label{lagoptwo}
\end{eqnarray}
where $L_{\mu}=-i f_{\pi}^2 U D_{\mu} U^{\dagger}$ with 
$U=\exp(-i\vec{\lambda}\cdot\vec{\phi}(x))/f_{\pi}$\cite{pichder91}.
The function $\vec{\phi}$ represents the octet of
pseudo-Goldstone bosons. 
The low-energy constants $g_{27}^{(1/2)}$ and $g_{27}^{(3/2)}$ are 
associated with $|\Delta I|=1/2$
and $3/2$ $(27_L,1_R)$ transitions, respectively. 
We retain $g_{27}^{(1/2)}$ and $g_{27}^{(3/2)}$ as distinct
entities as we anticipate the SU(3)$_f$ relation 
$g_{27}^{(1/2)}=g_{27}^{(3/2)}/5$ is broken at higher orders in the
weak chiral expansion --- we will see what other features 
are required to incorporate the effects of higher-order terms
in a systematic manner.
No ``weak mass'' term occurs in leading order 
in the weak chiral Lagrangian~\cite{cronin}, so that 
$m_u \ne m_d$ effects appear 
exclusively through $\pi^0$-$\eta$ mixing, as realized
in Eq.~(\ref{piphys}), and meson mass differences. 
In \optwo and to leading order in $(m_d-m_u)$, we have
\begin{eqnarray}
A_{K^0\ra\pi^+\pi^-}&=&
\sqrt{2}C i \left(g_8 + g_{27}^{(1/2)} + g_{27}^{(3/2)}
+ \frac{2\epsilon_8}{\sqrt{3}}(g_8 + g_{27}^{(1/2)} 
+ g_{27}^{(3/2)})\right) \non\\
A_{K^0\ra\pi^0\pi^0}&=&
\sqrt{2}C i \left(g_8 + g_{27}^{(1/2)} - 2g_{27}^{(3/2)}
- \frac{2\epsilon_8}{\sqrt{3}}(5 g_{27}^{(1/2)} - g_{27}^{(3/2)})\right) 
\label{paramotwo}\\
A_{K^+\ra\pi^+\pi^0}&=&
C i \left(3g_{27}^{(3/2)}
+ \frac{\epsilon_8}{\sqrt{3}}(2 g_8 + 12 g_{27}^{(1/2)} - 
3 g_{27}^{(3/2)})\right) \;, \non 
\end{eqnarray}
where $\epsilon_8=\sqrt{3}/4((m_d - m_u)/(m_s - \hat{m}))$ 
and $C=-(G_{F}/\sqrt{2})V_{ud} V_{us}^\ast f_{\pi} (m_s - \hat{m})B_0$,
and $(m_s - \hat{m})B_0=m_K^2-m_\pi^2$ in the isospin-perfect limit.
We thus recover 
\begin{eqnarray}
A_0 + \delta A_{1/2}&=&
C \left(\sqrt{2}(g_8 + g_{27}^{(1/2)}) + 
\frac{2}{3}
\sqrt{\frac{2}{3}}\epsilon_8
(2 g_8 - 3 g_{27}^{(1/2)} 
+ 3 g_{27}^{(3/2)})\right) \non\\
A_2 + \delta A_{3/2}&=&
C \left(2 g_{27}^{(3/2)} + 
\frac{2}{\sqrt{3}}\epsilon_8 (\frac{2}{3}g_8 + 4 g_{27}^{(1/2)}  
- \frac{3}{5} g_{27}^{(3/2)} ) \right) 
\label{loparam}\\
\delta A_{5/2}&=& \frac{2\sqrt{3}}{5} C \epsilon_8 g_{27}^{(3/2)} \non \;
\end{eqnarray}
and
\begin{equation}
x=\frac{\sqrt{2} r^{(3/2)} }{1 + r^{(1/2)}}\left(1 - 
\frac{2}{3\sqrt{3}}\epsilon_8 
\frac{(2 + 3(r^{(3/2)} - r^{(1/2)})}{1+ r^{(1/2)}}
 \right) + \frac{\epsilon_8}{15}\sqrt{\frac{2}{3}}
\frac{(10 - 9r^{(3/2)}  + 60 r^{(1/2)})}{1 + r^{(1/2)}}
\label{lox}
\end{equation}
\begin{equation}
y=\frac{\sqrt{6}}{5} \frac{\epsilon_8 r^{(3/2)}}{1 + r^{(1/2)}}\;,
\label{loy}
\end{equation}
where 
$r^{(1/2)}\equiv g_{27}^{(1/2)}/g_8$ and 
$r^{(3/2)}\equiv g_{27}^{(3/2)}/g_8$. 
We will allow $r^{(1/2)} \ne r^{(3/2)}/5$ 
in our fits as well, in order to ape the 
inclusion of higher-order effects
in the weak chiral Lagrangian.
Were the fits in the isospin-symmetric limit 
a reasonable estimate of the low-energy constants, so that
Eq.(\ref{R2xy}) yields $|A_2/A_0| \sim 0.045$\cite{devdic}, 
we would expect $|y|$ to be roughly $1.7 \cdot 10^{-4}$, 
as $(m_s - {\hat m})/(m_d - m_u)=40.8 \pm 3.2$~\cite{leut96}. 
This implies that we really must include electromagnetic effects
in our analysis as well. 
\begin{table}[htb]
\begin{center}
\caption{
The values of $r^{(1/2)}$ and $r^{(3/2)}$ 
determined by fitting Eqs.(\protect{\ref{loxem}})
and (\protect{\ref{loyem}}) to 
the empirically determined $x$ and $y$, 
resulting from the phase shift differences, $\delta_0-\delta_2$, 
compiled from various sources. 
We also show the values of $r^{(1/2)}$ and $r^{(3/2)}$ which
result were the central value of 
$\delta_0-\delta_2$ 1$\sigma$ or 2$\sigma$
larger than that reported by Ref.~\protect{\cite{meissner}}.
Electromagnetic effects are included as per 
Ref.~\protect{\cite{cirigliano}}. 
Note that (C) and (D) denote the 
results as computed in
chiral perturbation theory (C) and in 
the ``dispersive
matching'' (D) approach, respectively. We also show the
values of $r^{(1/2)}$ and $r^{(3/2)}$ which result
if the electromagnetically-generated phase shift, 
$\gamma_2\simeq 4.5^\circ$~\protect{\cite{cirigliano}},
is included --- using Ref.~\protect{\cite{meissner}}
this effectively implies $\delta_0-\delta_2=40.5^\circ$. 
The parameter $\epsilon_8$ is taken to be
$\epsilon_8=0.0106 \pm 0.0008\;$~\protect{\cite{leut96}}.
Solutions yielding $A_2/A_0 \gtrsim 1$ have been omitted.
For comparison, note that the 
analysis of $R_2$ in the isospin perfect limit yields 
$|x|=|A_2/A_0|= 0.0449\pm 0.0003$~\protect{\cite{devdic}}.
\smallskip
}
\begin{tabular}{cccc}
\hline
$\delta_0-\delta_2  = 40.5^\circ \pm 6^\circ $  & 
 $x=0.0394 \pm 0.0018$   & $y = -0.0082\pm 0.0027$ \\
\hline
$r^{(3/2)}$  & $r^{(1/2)}$ & $A_2/A_0$ & $\hbox{em}$ \\
 $   0.0173 \pm   0.0075 $ &
 $  -1.32 \pm   0.28 $ &
 $  -0.078 \pm   0.078 $ &  $\hbox{C}$ \\ 
 $   0.0176 \pm   0.0079 $ &
 $  -1.36 \pm   0.19 $ &
 $  -0.069 \pm   0.046 $ &  $\hbox{D}$ \\ 
\hline
$\delta_0-\delta_2  = 42^\circ \pm 4^\circ $ \protect{\cite{chell}} & 
 $x=0.0398 \pm 0.0016$   & $y = -0.0077\pm 0.0024$ \\
\hline
$r^{(3/2)}$  & $r^{(1/2)}$ & $A_2/A_0$ & $\hbox{em}$ \\
 $   0.0173 \pm   0.0077 $ &
 $  -1.34 \pm   0.30 $ &
 $  -0.073 \pm   0.074 $ & $\hbox{C}$ \\ 
 $   0.0175 \pm   0.0081 $ &
 $  -1.39 \pm   0.20 $ &
 $  -0.064 \pm   0.043 $ & $\hbox{D}$ \\
\hline
$\delta_0-\delta_2  = 45^\circ \pm 6^\circ $ \protect{\cite{meissner}} & 
 $x=0.0405 \pm 0.0021$ & $y = -0.0066\pm 0.0032$ \\
\hline
$r^{(3/2)}$  & $r^{(1/2)}$ & $A_2/A_0$ & $\hbox{em}$ \\
 $   0.0171 \pm   0.0083 $ &
 $  -1.39 \pm   0.37 $ &
 $  -0.061 \pm   0.069 $ & $\hbox{C}$ \\ 
 $   0.0172 \pm   0.0087 $ &
 $  -1.45 \pm   0.28 $ &
 $  -0.054 \pm   0.044 $ & $\hbox{D}$ \\ 
\hline
$\delta_0-\delta_2  = 51^\circ \pm 6^\circ $ & 
 $x=0.0424 \pm 0.0027$ & $ y=-0.0038 \pm 0.0041$ \\
\hline
$r^{(3/2)}$  & $r^{(1/2)}$ & $A_2/A_0$ & $\hbox{em}$ \\
 $   0.015 \pm   0.012 $ &
 $  -1.69 \pm   0.93 $ &
 $  -0.032 \pm   0.058 $ & $\hbox{C}$ \\ 
 $   0.015 \pm   0.013 $ &
 $  -1.79 \pm   0.91 $ &
 $  -0.027 \pm   0.045 $ & $\hbox{D}$ \\ 
\hline
\hline
$\delta_0-\delta_2  = 57^\circ \pm 6^\circ $ & 
$x=0.0451\pm  0.0037$ & $y= 0.00027 \pm 0.0055$ \\
\hline
$r^{(3/2)}$  & $r^{(1/2)}$ & $A_2/A_0$ & $\hbox{em}$ \\
 $   0.1 \pm   1.8 $ &
 $   10 \pm   250 $ &
 $   0.013 \pm   0.064 $ & $\hbox{C}$ \\ 
 $   0.1 \pm   2.1 $ &
 $   12 \pm   290 $ &
 $   0.013 \pm   0.057 $ & $\hbox{D}$ \\ 
\hline
\end{tabular}
\label{tableall}
\end{center}
\end{table}
The electromagnetically-induced phase shifts appear 
to be small~\cite{cirigliano}, so that we merely include the
modifications to the amplitudes themselves~\cite{neglectem2}. 
Following Ref.~\cite{cirigliano}, we have
\begin{eqnarray}
 \delta A_{1/2}^{em} &=&
\sqrt{2} C_{em} C g_8 \left(\frac{2}{3} C_{+-} + \frac{1}{3} C_{00} \right)
 \non\\
 \delta A_{3/2}^{em} 
&=&
\frac{2}{5} C_{em} C g_8 \left(\frac{2}{3} (C_{+-} - C_{00}) + C_{+0} 
\right) \\
\delta A_{5/2}^{em} &=& 
\frac{2}{5} C_{em} C g_8 \Big(C_{+-} - C_{00} -  C_{+0} 
\Big) 
\non \;
\end{eqnarray}
where $C_{em}=(f_{\pi}/f_K)(\alpha/4\pi)(1 + 2\hat{m}/(m_s - \hat{m}))$
and the ``dispersive matching'' approach of 
Ref.~\cite{cirigliano} yields
$C_{+-}=14.8 \pm 3.5$, $C_{00}=1.8 \pm 2.1$, and 
$C_{+0}=-7.1 \pm 7.4$. In the numerical estimates
we use $2{\hat m}/(m_s - \hat{m})= 
(m_{\pi^0}^2 + m_{\pi^+}^2)/(
m_{K^0}^2 + m_{K^+}^2 - (m_{\pi^0}^2 + m_{\pi^+}^2))$.
Only electromagnetic effects associated with 
$(8_L,1_R)$ transitions have been considered, as the
$|\Delta I|=1/2$ rule suggests they ought dominate. 
Including electromagnetic effects thus yields
\begin{eqnarray}
x&=&\frac{\sqrt{2} r^{(3/2)} }{1 + r^{(1/2)}}\left(1 - 
\frac{2}{3\sqrt{3}}\epsilon_8 
\frac{(2 + 3(r^{(3/2)} - r^{(1/2)})}{1+ r^{(1/2)}}
- 
\frac{C_{em} (2 C_{+-} + C_{00})}{3(1+ r^{(1/2)})}
 \right) 
\label{loxem} \\ 
&+& \frac{\epsilon_8}{15}\sqrt{\frac{2}{3}}
\frac{(10 - 9r^{(3/2)}  + 60 r^{(1/2)})}{1 + r^{(1/2)}}
+ \frac{\sqrt{2}}{5}
\frac{C_{em} (2(C_{+-} - C_{00}) + 3C_{+0})}{3(1+ r^{(1/2)})}
\non
\end{eqnarray}
and
\begin{equation}
y=\frac{\sqrt{2}}{5}\left( \frac{\sqrt{3} \epsilon_8 r^{(3/2)} 
+ C_{em}(C_{+-} - C_{00} - C_{+0})}
{1 + r^{(1/2)}}\right)\;. 
\label{loyem}
\end{equation}
Using Ref.~\cite{cirigliano} we have 
$C_{em}(C_{+-} - C_{00} - C_{+0})=0.0029 \pm 0.0019$, as 
$f_K/f_{\pi} = 1.23 \pm 0.02$~\cite{DGH}. Consequently
if $r^{(3/2)}$ were as small as the isospin-symmetric
limit would imply, then $y$ ought be given by  
$C_{em}(C_{+-} - C_{00} - C_{+0})$, yet they are of opposite
sign. This implies that the error in the $\delta_0 - \delta_2$
phase shift is even larger, or that the errors in the calculations
of the electromagnetic effects are underestimated. 
Nevertheless, as apparently $y$ is negative and 
$C_{em}(C_{+-} - C_{00} - C_{+0})$ is positive, the discrepancy
could be resolved by adjusting $r^{(1/2)}$ and $r^{(3/2)}$ 
to suit the empirically determined $x$ and $y$. Let us
examine this point explicitly. In Table~\ref{tableall} we
show the values of $r^{(1/2)}$ and $r^{(3/2)}$ which emerge from
fitting the values of $x$ and $y$ which result from
the empirical branching ratios and various values of the
$\delta_0-\delta_2$ phase shift difference.

The salient points of our analysis can be summarized as follows. 
\begin{itemize}

\item If the SU(3)$_f$ relation $r^{(1/2)}=r^{(3/2)}/5$ 
is imposed, then the value of $\epsilon_8$ which
emerges is ${\cal O}(20\%)$ and is thus untenably large. 

\item If the SU(3)$_f$ relation $r^{(1/2)}=r^{(3/2)}/5$ 
is no longer imposed, and $\epsilon_8$ is fixed as per 
$\epsilon_8=0.0106 \pm 0.0008$~\protect{\cite{leut96}},
then $r^{(1/2)}$ is {\it very} different from $r^{(3/2)}/5$
--- the SU(3)$_f$ breaking effects are extremely large. 
This result is driven by large difference between 
the empirical value of $y$ 
and the computed electromagnetic contribution~\cite{cirigliano}. 
That is, if we were to drop terms of 
${\cal O}(r^{(3/2)}\epsilon_8)$ all together, 
then, with 
$\delta_0-\delta_2=45^\circ$, 
 Eq.~(\ref{loyem}) implies that $r^{(1/2)}=-1.440$ and 
Eq.~(\ref{loxem}) implies that $r^{(3/2)}=0.0184$.
The inclusion of ${\cal O}(r^{(3/2)}\epsilon_8)$ terms
do not significantly reduce this difficulty. 
Such large SU(3)$_f$ breaking effects are difficult
to reconcile with chiral power counting and 
model estimates, which suggest
such effects are no more than  30\%~\cite{pich86}. 

\item The value of $A_2/A_0$ is generally different from and
rather more uncertain than
that which emerges from Eq.~(\ref{R2xy}) in the
isospin-symmetric limit, namely 
$|A_2/A_0|\approx 0.045$ with $A_2/A_0 >0$. 

\end{itemize}

The breaking of SU(3)$_f$ relation $r^{(1/2)}=r^{(3/2)}/5$ 
apes the inclusion of higher order effects in the
weak chiral Lagrangian, and the large breaking effects
seen suggest that including \opfour effects are
very important. This has some precedent, as 
in the isospin-symmetric limit, 
Ref.~\cite{kmwplb} finds a $30\%$ quenching of the \optwo $g_8$ 
result in \opfour. The SU(3)$_f$ breaking effects 
seen, however, are much too large~\cite{pich86}
and prompt an investigation of the presence of higher-order effects 
in a more systematic fashion. 

We wish to consider how \opfour effects impact
the parametrization of Eq.~(\ref{paramotwo}). 
We enlarge our parametrization by considering how
the terms of the \opfour 
weak, chiral Lagrangian of Ref.~\cite{kambor}
may be reorganized into the form of Eq.~(\ref{loparam}). 
We distinguish the ${\cal O}(m_d - m_u)$ terms which arise
from ``kinematics,'' i.e., from factors of $m_{K^0}$,
from $\pi^0$-$\eta$ mixing, as well as from the counterterms
of the ${\cal O}(p^4)$, weak, chiral Lagrangian. 
We find that the effects of the higher-order terms can
be absorbed in this case into {\it effective} $g_8$, 
$g_{27}^{(1/2)}$, and $g_{27}^{(3/2)}$ constants, with
one additional phenomenological amplitude 
$\delta \tilde A_{5/2}^{\rm h.o.}$, generated by 
\opfour contributions of $(27_L,1_R)$ character times
$B_0(m_d - m_u)$. 
Varying the possible inputs within
the bounds suggested by dimensional analysis, we are
unable to reduce the SU(3)$_f$ breaking of the 
relation $r^{(1/2)}=r^{(3/2)}/5$ to the level 
needed if the additional phenomenological
$\delta \tilde A_{5/2}^{\rm h.o.}$ is generated solely
by $m_d\ne m_u$ effects. 
Thus we are unable to construct
a suitable phenomenological description of the $K\ra\pi\pi$
amplitudes with the 
$\delta_0-\delta_2$ phase
shift of Ref.~\cite{meissner}
and with the computed electromagnetic effects of
Ref.~\cite{cirigliano}.
The size of $\delta \tilde A_{5/2}^{\rm h.o.}$ required
to generate suitably small violations 
of $r^{(1/2)}=r^{(3/2)}/5$ 
suggests the presence of missing electromagnetic 
effects generated by $(8_L,1_R)$ operators. 
The authors of Ref.~\cite{cirigliano}
are in the process of estimating additional electromagnetic 
effects~\cite{jfdcomm}.
The details of our efforts are delineated in the Appendix. 
Note that issues of a similar ilk have been addressed in Ref.~\cite{wolfe}. 

It is worth noting that the conundrum we have been unable
to resolve is 
unlikely to be due to ``new'' physics in 
$K\ra\pi\pi$ decays. The operator-product expansion
for $s\ra d {\overline q} q$ transitions starts in dimension six,
so that at most three $u,d$ quark fields are present, implying
that the short-distance operators generate at most a 
$|\Delta I|=3/2$ transition. In next-to-leading order, 
as many as five $u,d$ quark fields are present, so that
a short-distance $|\Delta I|=5/2$ transition is possible. 
Thus to estimate the plausibility of physics beyond
the Standard Model as a source of $|\Delta I|=5/2$ effects, 
we need only estimate the relative importance of dimension-nine 
to dimension-six operators.
Each new dimension is suppressed by the scale $\Lambda$ ---
in the Standard Model, $\Lambda \sim M_W$, otherwise 
$\Lambda > M_W$. For $K\ra\pi\pi$ decays the relative 
importance of the dimension-nine operators is no larger
than $(M_K/M_W)^3$. Clearly
short-distance physics cannot generate an appreciable
$|\Delta I|=5/2$ amplitude, so that the presence of physics
beyond the Standard Model cannot be invoked to reconcile
our difficulty. 

The presence of a $|\Delta I| =5/2$ amplitude also impacts the 
theoretical value of \epsrat, for standard practice employs
a value of $\omega$ determined from the $K\ra \pi\pi$ branching
ratios under the assumption that isospin symmetry is perfect.
We proceed to investigate how the presence of a 
$|\Delta I|=5/2$ amplitude impacts the value of \epsrat.

\section{Isospin Violation in 
Re$\,(\epsilon^\prime/\epsilon)$}
\label{isoeps}

We wish to examine how isospin-violating effects
impact the theoretical value of \reeps and the extraction of the
value of $\omega$, namely the ratio ${\rm Re}\,A_2/{\rm Re}\,A_0$, 
where $A_I$ denotes
the amplitude for $K\ra (\pi\pi)_I$ and $(\pi\pi)_I$ denotes
a $\pi\pi$ final state of isospin $I$.
The empirical value of \reeps is inferred from the following
ratio of ratios~\cite{ktev,oldeps}:
\begin{equation}
{\rm Re}\,(\frac{\epsilon^\prime}{\epsilon}) = \frac{1}{6}
\left[
\Bigg|\frac {\eta_{+-}}{\eta_{00}}
\Bigg|^2 -1
\right]\;, 
\label{epsdefexp}
\end{equation}
where 
\begin{equation}
\eta_{+-} \equiv 
\frac{\langle \pi^+ \pi^- | {\cal H}_W | K_L^0\rangle}
{\langle \pi^+ \pi^- | {\cal H}_W | K_S^0\rangle}
\quad;\quad
\eta_{00} \equiv 
\frac{\langle \pi^0 \pi^0 | {\cal H}_W | K_L^0\rangle}
{\langle \pi^0 \pi^0 | {\cal H}_W | K_S^0\rangle} \;
\end{equation}
and ${\cal H}_W$ is the effective weak Hamiltonian for kaon decays. 
Writing $K_S^0$ and $K_L^0$ in terms of the CP eigenstates
$|K_{\pm}^0\rangle$ yields 
$|K_{L,S}^0\rangle =
(|K_{\mp}^0\rangle + {\overline \vep}|K_{\pm}^0\rangle)/
\sqrt{1 + |{\overline\vep}|^2}$, 
noting that $|K_{\pm}^0\rangle =
(|K^0\rangle \mp |{\overline K}^0\rangle)/\sqrt{2}$. 
Using Eq.~(\ref{trueparam}) and 
treating the weak phases 
as small, 
so that only leading-order terms in 
$\xi_0,\xi_2, \xi_{00}$, and $\xi_{+-}$ are retained, we find
\begin{equation}
\eta_{+-} =\epsilon
+ i \frac{ 
\frac{1}{\sqrt{2}} 
| \frac{A_2}{A_0}| (\xi_2 - \xi_0)
e^{i(\delta_2 - \delta_0)}
+ | \frac{A_{\rm IB}^{+-}}{A_0}| (\xi_{+-} - \xi_0)
e^{i(\delta_{+-} - \delta_0)}
}{1 + 
\frac{1}{\sqrt{2}} | \frac{A_2}{A_0}| 
e^{i(\delta_2 - \delta_0)}
+ | \frac{A_{\rm IB}^{+-}}{A_0}| 
e^{i(\delta_{+-} - \delta_0)}
}
\label{etapm}
\end{equation}
and 
\begin{equation}
\eta_{00} =\epsilon
- i\frac{ 
\sqrt{2} | \frac{A_2}{A_0}| (\xi_2 - \xi_0)
e^{i(\delta_2 - \delta_0)}
-  
| \frac{A_{\rm IB}^{00}}{A_0}| (\xi_{00} - \xi_0)
e^{i(\delta_{00} - \delta_0)}
}{1 
- \sqrt{2} | \frac{A_2}{A_0}| 
e^{i(\delta_2 - \delta_0)} 
+ | \frac{A_{\rm IB}^{00}}{A_0}| 
e^{i(\delta_{00} - \delta_0)}
}\;, 
\label{eta00}
\end{equation}
where $\epsilon\equiv\bar{\vep} + i \xi_0$. 
Defining 
\begin{equation}
\frac{\eta_{+-}}{\eta_{00}}
\equiv 1 + 3 \frac{\epsilon^\prime}{\epsilon}
\label{epsdefth}
\end{equation}
and retaining the leading terms in 
$|A_{\rm IB}^{+-}/A_0|$, $|A_{\rm IB}^{00}/A_0|$, and weak phases, we
have 
\begin{eqnarray}
\frac{\epsilon^\prime}{\epsilon}
&=&
\frac{i e^{i(\delta_2 -\delta_0 - \Phi_{\epsilon})}}{\sqrt{2}|\epsilon|}
\Bigg[\bigg|\frac{A_2}{A_0}\bigg|(\xi_2 - \xi_0) 
\left[1 + \frac{1}{\sqrt{2}}e^{i(\delta_2 - \delta_0)}
\bigg|\frac{A_2}{A_0}\bigg|
- \frac{1}{3}
\left( e^{i(\delta_{+-} - \delta_0)}\bigg|\frac{A_{\rm IB}^{+-}}{A_0}\bigg|
+ 2
e^{i(\delta_{00} - \delta_0)}\bigg|\frac{A_{\rm IB}^{00}}{A_0}\bigg|
\right)
\right] \non\\
&& \qquad \quad +
\frac{\sqrt{2}}{3}\left[
e^{i(\delta_{+-} - \delta_2)}
\bigg|\frac{A_{\rm IB}^{+-}}{A_0}\bigg|(\xi_{+-} - \xi_0) 
-
e^{i(\delta_{00} - \delta_2)}
\bigg|\frac{A_{\rm IB}^{00}}{A_0}\bigg|(\xi_{00} - \xi_0) 
\right] 
\label{epsdef1} \\
&& \qquad \quad - 
\frac{1}{3}\frac{|A_2|}{|A_0|}
\left[
e^{i(\delta_{+-} - \delta_0)}
\bigg|\frac{A_{\rm IB}^{+-}}{A_0}\bigg|(\xi_{+-} - \xi_0) 
+ 2
e^{i(\delta_{00} - \delta_0)}
\bigg|\frac{A_{\rm IB}^{00}}{A_0}\bigg|(\xi_{00} - \xi_0) 
\right]
\Bigg] \;,\non 
\end{eqnarray}
where we have retained terms of ${\cal O}(|A_2/A_0|^2)$ as
well, for consistency. 
Note that $\epsilon=|\epsilon|e^{i\Phi_{\epsilon}}$. 
Equation (\ref{epsdefth}) is consistent with the empirical definition
of Eq.~(\ref{epsdefexp}) as corrections 
of $(\epsilon^\prime/\epsilon)^2$ are trivial. 
Alternatively, 
\begin{eqnarray}
\frac{\epsilon^\prime}{\epsilon}
&=&
-
\frac{i \xi_0\omega e^{i(\delta_2 -\delta_0 - \Phi_{\epsilon})}}
{\sqrt{2}|\epsilon|}
\Bigg(
1 - \frac{1}{\omega}
\Bigg(
\bigg|\frac{A_2}{A_0}\bigg|
\frac{\xi_2}{\xi_0} 
\left[ 1 + \frac{1}{\sqrt{2}} e^{i(\delta_2 - \delta_0)}
\bigg|\frac{A_2}{A_0}\bigg|
- \frac{1}{3}
\left[ e^{i(\delta_{+-} - \delta_0)}\bigg|\frac{A_{\rm IB}^{+-}}{A_0}\bigg|
+ 2
e^{i(\delta_{00} - \delta_0)}\bigg|\frac{A_{\rm IB}^{00}}{A_0}\bigg|
\right]
\right] \non \\
&& \qquad +
\frac{\sqrt{2}}{3}\left[
e^{i(\delta_{+-} - \delta_2)}\bigg|\frac{A_{\rm IB}^{+-}}{A_0}\bigg|
\frac{\xi_{+-}}{\xi_0}
-
e^{i(\delta_{00} - \delta_2)}\bigg|\frac{A_{\rm IB}^{00}}{A_0}\bigg|
\frac{\xi_{00}}{\xi_0} 
\right]     
\label{epsdef2} \\
&& \qquad - 
\frac{1}{3}\frac{|A_2|}{|A_0|}
\left[
e^{i(\delta_{+-} - \delta_0)}\bigg|\frac{A_{\rm IB}^{+-}}{A_0}\bigg|
\frac{\xi_{+-}}{\xi_0}
+ 2
e^{i(\delta_{00} - \delta_0)}
\bigg|\frac{A_{\rm IB}^{00}}{A_0}\bigg|
\frac{\xi_{00}}{\xi_0}
\right]
\Bigg)\Bigg)\;, \non
\end{eqnarray}
where
\begin{eqnarray}
\omega &=&
\bigg|\frac{A_2}{A_0}\bigg| +
\frac{\sqrt{2}}{3}\left(
e^{i(\delta_{+-} - \delta_2)}\bigg|\frac{A_{\rm IB}^{+-}}{A_0}\bigg|
-
e^{i(\delta_{00} - \delta_2)}\bigg|\frac{A_{\rm IB}^{00}}{A_0}\bigg|
\right)
+ \frac{1}{\sqrt{2}} e^{i(\delta_2 - \delta_0)}\bigg|\frac{A_2}{A_0}\bigg|^2 
\\
&-& 
\frac{2}{3}\bigg|\frac{A_2}{A_0}\bigg| 
\left(
e^{i(\delta_{+-} - \delta_0)}\bigg|\frac{A_{\rm IB}^{+-}}{A_0}\bigg|
+ 2
e^{i(\delta_{00} - \delta_0)}
\bigg|\frac{A_{\rm IB}^{00}}{A_0}\bigg|
\right) \;. \non
\label{omegadef}
\end{eqnarray}
Thus, working to leading order in isospin violation
and ignoring electromagnetic effects in the ``strong'' phases, 
specifically implying as per Eqs.~(\ref{trueparam}) and (\ref{newparam})
that
\begin{eqnarray}
A_{\rm IB}^{+-} e^{i\delta_{+-}}
&=& \delta A_{1/2} e^{i\delta_0}
+ \frac{1}{\sqrt{2}} 
\left( \delta A_{3/2} + \delta A_{5/2} \right) 
e^{i\delta_2} 
\non \\
A_{\rm IB}^{00} e^{i\delta_{00}}
&=& \delta A_{1/2} e^{i\delta_0}
- \sqrt{2} 
\left( \delta A_{3/2} + \delta A_{5/2} \right)
e^{i\delta_2} 
\label{paramrel} \;,
\end{eqnarray}
Eqs.~(\ref{epsdef2},\ref{omegadef}) become
\begin{eqnarray}
\frac{\epsilon^\prime}{\epsilon} = -
\frac{i \xi_0\omega e^{i(\delta_2 -\delta_0 - \Phi_{\epsilon})}}
{\sqrt{2}|\epsilon|}
\Bigg(1 &-& \frac{1}{\omega}\Bigg(
\bigg|\frac{A_2}{A_0}\bigg|\frac{\xi_2}{\xi_0} + 
\frac{1}{\sqrt{2}}e^{i(\delta_2 - \delta_0)}
 \bigg|\frac{A_2}{A_0}\bigg|^2 \frac{\xi_2}{\xi_0} + 
\frac{{\rm Im}(\delta A_{3/2} + \delta A_{5/2})}{|A_0|\xi_0} 
\non \\
&-&
\bigg|\frac{A_2}{A_0}\bigg|\frac{\xi_2}{\xi_0} 
\left[ 
\frac{{\rm Re}\, \delta A_{1/2}}{|A_0|} 
- \frac{1}{\sqrt{2}} e^{i(\delta_2 - \delta_0)}
\frac{{\rm Re}(\delta A_{3/2} + \delta A_{5/2})}{|A_0|}
\right] 
\label{epsdef2nn} \\
&-&
\bigg|\frac{A_2}{A_0}\bigg|
\left[ 
\frac{{\rm Im}\, \delta A_{1/2}}{|A_0| \xi_0} 
- \frac{1}{\sqrt{2}} e^{i(\delta_2 - \delta_0)}
\frac{{\rm Im}(\delta A_{3/2} + \delta A_{5/2})}{|A_0| \xi_0}
\right]
\Bigg)\Bigg)\;, \non
\end{eqnarray}
where
\begin{eqnarray}
\omega &=&
\bigg|\frac{A_2}{A_0}\bigg| + 
\frac{{\rm Re}(\delta A_{3/2} + \delta A_{5/2})}{|A_0|} 
+ 
\frac{1}{\sqrt{2}}e^{i(\delta_2 - \delta_0)}
\bigg|\frac{A_2}{A_0}\bigg|^2  + 
\non \\
&-&
2 \bigg|\frac{A_2}{A_0}\bigg| 
\left[ 
\frac{{\rm Re}\, \delta A_{1/2}}{|A_0|} 
- \frac{1}{\sqrt{2}} e^{i(\delta_2 - \delta_0)}
\frac{{\rm Re}(\delta A_{3/2} + \delta A_{5/2})}{|A_0|}
\right] \;.
\label{omegadefnn}
\end{eqnarray}
We can recast these formulae into a more familiar form~\cite{buraslh} by
writing Eq.~(\ref{epsdef2nn}) as 
\begin{equation}
\frac{\epsilon^\prime}{\epsilon}
= - \frac{i\omega e^{i(\delta_2 - \delta_0-\Phi_{\epsilon})}}
{\sqrt{2}|\epsilon|{\rm Re}\,A_0}
\left\{
{\rm Im}\, A_0 (1 - \Omega_{\rm IB})
- \frac{1}{\omega} {\rm Im}\, A_2
- \frac{1}{\sqrt{2}}e^{i(\delta_2 -\delta_0)} {\rm Im}\, A_2
\right\} \,,
\label{epsburasplus}
\end{equation}
where $\omega$ is defined by Eq.~(\ref{omegadefnn}) and 
\begin{eqnarray}
\Omega_{\rm IB}
&=& \frac{1}{\omega}
\Bigg(
\frac{{\rm Im}(\delta A_{3/2} + \delta A_{5/2})}{{\rm Im} A_0} 
-
\frac{{\rm Im} A_2}{{\rm Im} A_0}
\left[ 
\frac{{\rm Re}\, \delta A_{1/2}}{|A_0|} 
- \frac{1}{\sqrt{2}} e^{i(\delta_2 - \delta_0)}
\frac{{\rm Re}(\delta A_{3/2} + \delta A_{5/2})}{|A_0|}
\right] 
\label{Omegaib} \\
&-&
\bigg|\frac{A_2}{A_0}\bigg|
\left[ 
\frac{{\rm Im}\, \delta A_{1/2}}{{\rm Im} A_0} 
- \frac{1}{\sqrt{2}} e^{i(\delta_2 - \delta_0)}
\frac{{\rm Im}(\delta A_{3/2} + \delta A_{5/2})}{{\rm Im} A_0} 
\right]
\Bigg)
\;. \non
\end{eqnarray}
If we assume that the $|\Delta I|=1/2$ 
enhancement observed in ${\rm Re}\,A_I$ is
germane to ${\rm Im}\,A_I$ as well, so that 
both ${\rm Re}\, A_0 \gg {\rm Re}\, A_2$
and ${\rm Im}\, A_0 \gg {\rm Im}\, A_2$ are
satisfied, then if we ignore terms of 
${\cal O}(({\rm Re}\, A_2/{\rm Re}\, A_0)(\epsilon_8,\alpha))$
and of 
${\cal O}(({\rm Im}\, A_2/{\rm Im}\, A_0)(\epsilon_8,\alpha))$,
as well as of ${\cal O}((|A_2|/|A_0|)^2)$,
we find that Eq.~(\ref{epsburasplus}) can be written as~\cite{buraslh}
\begin{equation}
\frac{\epsilon^\prime}{\epsilon}
= - \frac{i\omega e^{i(\delta_2 - \delta_0-\Phi_{\epsilon})}}
{\sqrt{2}|\epsilon|{\rm Re}\,A_0}
\left\{
{\rm Im}\, A_0 (1 - \Omega_{\rm IB})
- \frac{1}{\omega} {\rm Im}\, A_2
\right\} \,,
\label{epsburas}
\end{equation}
with 
\begin{equation}
\Omega_{\rm IB}
= \frac{1}{\omega}
\left(
\frac{{\rm Im}(\delta A_{3/2} + \delta A_{5/2})}{{\rm Im} A_0}
\right)
\label{simpleomegaib}
\end{equation}
and
\begin{equation}
\omega =
\bigg|\frac{A_2}{A_0}\bigg| + 
\frac{{\rm Re}(\delta A_{3/2} + \delta A_{5/2})}{|A_0|} \;.
\label{simpleomega}
\end{equation}
Equation (\ref{simpleomegaib}) is proportional to 
${\rm Im}A_{K^0 \ra \pi^+\pi^-} - {\rm Im}A_{K^0 \ra \pi^0\pi^0}$
and is generated by $(8_L,1_R)$ operators.
It is equivalent to Eq.(4) in Ref.~\cite{sggv}.

In standard practice, 
the value of $\omega$ is typically extracted from
the analysis of $K\ra\pi\pi$ branching ratios in the isospin-perfect
limit; specifically, $\omega$ is
set equal to the RHS of Eq.~(\ref{R2xy}), yielding~\cite{pdg98}
\begin{equation}
2\sqrt{\frac{R_2}{3}} \equiv \omega_{\rm exp} = 0.0449 \pm 0.0003 \;.
\end{equation}
From Eqs.~(\ref{R2xy},\ref{R1xy}), we see 
that $\omega$ as defined by Eq.~(\ref{omegadefnn}) is actually
given by 
\begin{equation}
\omega=\omega_{\rm exp} + \frac{5}{3} y + \frac{1}{\sqrt{2}} (x+y)^2
- 2  \frac{A_2}{A_0} 
\frac{{\rm Re}\,\delta A_{1/2}}{A_0} \;, 
\label{omegaeps}
\end{equation}
where we ignore terms of non-leading order in isospin violation, 
as well as terms of ${\cal O}((|A_2|/|A_0|)^2,(\alpha,\epsilon_8))$. 
If $\delta_0 -\delta_2 = 45^\circ \pm 6^\circ$~\cite{meissner,ananth}, 
then we find from Table \ref{tablexy} that the second term
of Eq.~(\ref{omegaeps}) is 
$\sim -0.0110$, whereas the third
term is $\sim 0.0008$. 
We estimate the last term of Eq.~(\ref{omegaeps}) 
to be $\sim \pm 2(0.045)(0.01) \sim \pm 0.0010$. 
Thus the last two terms are small relative to the error
in $y$ --- dropping them all together, we find~\cite{branco}
\begin{equation}
\omega = 0.0339 \pm 0.0056 \;.
\label{omeganum}
\end{equation}
The use of the value of 
$\omega$ given in Eq.~(\ref{omeganum}) tends to decrease
the SM prediction of \epsrat, both by an overall factor of
$\sim 25\%$, as well as by enhancing the cancellation of
the ${\rm Im}\, A_0$ and ${\rm Im}\, A_2$ contributions
of Eq.~(\ref{epsburas}). Note that our explicit estimate
of the additional terms included in Eq.~(\ref{omegadefnn})
suggests that the formulae of Eqs.~(\ref{epsburas}), (\ref{simpleomegaib}),
and (\ref{simpleomega}) characterize the isospin-violating
contributions in a sufficiently accurate manner. 
In order to assess the impact of
our numerical estimate of Eq.~(\ref{omeganum}), let us
turn to the schematic formula~\cite{burasr}
\begin{equation}
\frac{\epsilon^\prime}{\epsilon}
= 13 \,{\rm Im} \lambda_t \left[ B_6^{(1/2)}(1 - \Omega_{\eta+\eta^\prime})
- 0.4\, B_8^{(3/2)} 
\right] \;, 
\label{epsschem}
\end{equation}
in which $B_6^{(1/2)}=1.0$, $B_8^{(3/2)}=0.8$, 
${\rm Im} \lambda_t=1.3\cdot 10^{-4}$, 
$\Omega_{\eta+\eta^\prime}=0.25$, and $\omega= 0.045$
yields the  ``central'' SM value of 
$\epsilon^\prime/\epsilon
\sim 7 \cdot 10^{-4}$~\cite{burasr}. 
Using Eq.~(\ref{omeganum}) yields $\epsilon^\prime/\epsilon
\sim 4 \cdot 10^{-4}$, a 40\% decrease. It has been 
recently suggested that isospin-breaking effects
in the hadronization of the gluonic penguin operator
can generate isospin-breaking contributions to 
$\Omega_{\eta+\eta^\prime}$ beyond \pieta mixing, hence
$\Omega_{\eta+\eta^\prime}\ra \Omega_{\rm IB}$~\cite{sggv}. 
Interestingly, the use of the correct value of 
$\omega$, Eq.~(\ref{omeganum}), 
partially offsets the large increase in 
\epsrat found in Ref.~\cite{sggv}. 
Using the estimate $\Omega_{\rm IB} \ra -0.05 \ra -0.78$~\cite{sggv,epscomm}, 
based exclusively on strong-interaction isospin breaking, 
we find with Eqs.~(\ref{omeganum}) and (\ref{epsschem}) 
that 
\begin{equation}
\frac{\epsilon^\prime}{\epsilon} \sim (8 - 17) \cdot 10^{-4} 
\end{equation}
rather than 
\begin{equation}
\frac{\epsilon^\prime}{\epsilon} \sim (12 - 25) \cdot 10^{-4} 
\end{equation}
with $\omega=0.045$ and Eq.~(\ref{epsschem}). 
We anticipate that electromagnetic
effects also contribute to $\Omega_{\rm IB}$, so that 
our numerical estimates are certainly incomplete, though
indicative of the irreducible uncertainties present. 

It is useful to contrast the relations we have found for
\epsrat, $\omega$, and $\Omega_{\rm IB}$ 
with those used previously. 
Earlier treatments of strong-interaction isospin 
violation~\cite{bijwi,dght,buge} considered \pieta mixing exclusively,
as this is the only manner in which 
relevant $m_u \ne m_d$ effects appear in the 
O($p^2,1/N_c$) weak chiral Lagrangian. 
The $\eta^\prime$ enters as an explicit degree of freedom in
these treatments~\cite{dght,buge}. The small value of $\omega_{\rm exp}$ 
suggests that $(8_L,1_R)$ operators dominate the isospin-violating
contributions as well, and isospin violation based on the
$(27_L,1_R)$ contributions is thus neglected entirely.  
Assuming $(8_L,1_R)$ operators dominate the isospin-violating effects
means implicitly that the terms of 
${\cal O}(({\rm Re}\, A_2/{\rm Re}\, A_0)(\epsilon_8,\alpha))$
and of 
${\cal O}(({\rm Im}\, A_2/{\rm Im}\, A_0)(\epsilon_8,\alpha))$,
as well as of ${\cal O}((|A_2|/|A_0|)^2)$, are all neglected.
In the notation of Eq.~(\ref{trueparam}) 
$A_{\rm IB}^{+-}=0$ and $A_{\rm IB}^{00}=2(\varepsilon_{\eta}
\langle \pi^0 \eta | {\cal H}_W^8| K^0 \rangle 
+ \varepsilon_{\eta^\prime}
\langle \pi^0 \eta^\prime | {\cal H}_W^8 | K^0\rangle)$, where
$\varepsilon_{\eta}, \varepsilon_{\eta^\prime} \propto (m_d - m_u)$
and ${\cal L}_W^8$ 
denotes the effective weak Lagrangian transforming
as $(8_L,1_R)$ under ${\rm U(3)}_L\times{\rm U(3)}_R$ symmetry 
--- ${\cal L}_W^8$ contains exactly one term.
In Refs.~\cite{dght,buge}, the \pieta mixing contribution 
is incorporated by defining new I=0 and I=2 amplitudes,
such that the form of the isospin decomposition of Eq.~(\ref{isodecomp}) is
retained. Introducing $\Delta A_{0,2}\equiv A_{0,2} - A_{0,2}^{(0)}$
to describe the change in the $I=0$ and $I=2$ amplitudes under this
procedure we find 
\begin{equation}
\Delta A_2 = -\frac{\sqrt{2}}{3} A_{\rm IB}^{00} \quad ; \quad
\Delta A_0 = \frac{1}{3} A_{\rm IB}^{00}
\;. 
\label{deltaa}
\end{equation}
Thus one recovers the {\it form} of
Eq.~(\ref{epsburas}) with $\delta A_{3/2}=\delta A_{5/2}=0$. 
Rewriting the imaginary parts in terms of the isospin-perfect
pieces ${\rm Im}\,A_I^{(0)}$, i.e., in the absence 
of \pieta mixing, yields~\cite{buraslh} 
\begin{equation}
\frac{\epsilon^\prime}{\epsilon}
= - \frac{ie^{i(\delta_2 - \delta_0-\Phi_{\epsilon})}}
{\sqrt{2}|\epsilon|{\rm Re}\,A_0}
\left\{
\omega\,
{\rm Im}\, A_0^{(0)} (1 - \Omega_{\eta + \eta'})
- {\rm Im}\, A_2^{(0)}\right\}
\,
\label{epsdefold}
\end{equation}
with
\begin{eqnarray}
\Omega_{\eta + \eta'}&&=
\frac{1}{{\rm Im}\, A_0^{(0)}}
\left(\frac{{\rm Im}\,\Delta A_2}{\omega} - 
{\rm Im}\,\Delta A_0
\right) \non \\
&& \simeq
\frac{1}{\omega}
\frac{{\rm Im}\,\Delta A_2}{{\rm Im}\, A_0^{(0)}}  
\;,
\label{omeetadef}
\end{eqnarray}
noting that 
only the $\Delta A_2$ term is retained for phenomenological
purposes~\cite{buraslh}. 
Equation (\ref{epsdefold}) results from absorbing the isospin-violating
contributions into two amplitudes, ``$A_0$'' and ``$A_2$''. 
A third amplitude
is permitted in the presence of isospin violation. However, if we
neglect electromagnetic effects and 
consider isospin violation based on $(8_L,1_R)$ operators only, then
only two amplitudes are present, and the above 
procedure is appropriate. Equation~(\ref{epsdef2nn}) 
requires no such assumptions and thus is more
general than the expression in Eq.~(\ref{epsdefold}). 
Let us now consider Eq.~(\ref{epsburas}) in the event 
\pieta mixing were the only source of isospin-violation present --- 
we will continue to  assume that 
$(8_L,1_R)$ transitions generate the only numerically
important isospin-violating effects. 
Note that the ``kinematic'' $m_d\ne m_u$ effect
from $m_{K^0}^2$ does not contribute to 
$\delta A_{3/2} + \delta A_{5/2}$ in this case.
The mixing parameters $\epsilon_{\eta}$ and
$\epsilon_{\eta^\prime}$ are real~\cite{gl}, so that 
${\rm Im}(\delta A_{3/2} + \delta A_{5/2})$ is
determined by $\langle \eta^{(\prime)} | {\cal L}_W^8 | K^0 \rangle$.
The Lagrangian ${\cal L}_W^8$ contains exactly one term,
so that the matrix elements 
are proportional to $A_0^{(0)}$, and the
proportionality constant is real. Thus 
${\rm Im} (\delta A_{3/2} + \delta A_{5/2})/{\rm Im}\, A_0
= {\rm Re} (\delta A_{3/2} + \delta A_{5/2})/{\rm Re}\, A_0$
as \pieta mixing is real~\cite{gl},  so that we have 
\begin{equation}
\frac{\epsilon^\prime}{\epsilon}
=
-
\frac{i \xi_0\omega e^{i(\delta_2 -\delta_0 - \Phi_{\epsilon})}}
{\sqrt{2}|\epsilon|}
\left(1 - \frac{1}{\omega}\left(
|\frac{A_2}{A_0}|\frac{\xi_2}{\xi_0} -
\frac{\sqrt{2}}{3}
|\frac{A_{\rm IB}^{00}}{A_0}|
\right)\right)\;.
\label{epsp2}
\end{equation}
Using Eq.~(\ref{simpleomega}) we find 
\begin{eqnarray}
\frac{\epsilon^\prime}{\epsilon}
&=&
-
\frac{i \xi_0 e^{i(\delta_2 -\delta_0 - \Phi_{\epsilon})}}
{\sqrt{2}|\epsilon|}
\left( 
|\frac{A_2}{A_0}| - 
\frac{\sqrt{2}}{3}
|\frac{A_{\rm IB}^{00}}{A_0}|
- 
\left(
|\frac{A_2}{A_0}|\frac{\xi_2}{\xi_0} -
\frac{\sqrt{2}}{3}
|\frac{A_{\rm IB}^{00}}{A_0}|
\right)\right) \\
&=& 
-
\frac{i \xi_0 e^{i(\delta_2 -\delta_0 - \Phi_{\epsilon})}}
{\sqrt{2}|\epsilon|} 
|\frac{A_2}{A_0}| 
\left( 1 - 
\frac{\xi_2}{\xi_0} 
\right) 
\end{eqnarray}
and thus the inclusion of isospin-violating effects in 
\optwo acts to correct for isospin violation in the
extraction of $\omega$ from $K\ra\pi\pi$ branching ratios, 
to recover the ``true'' $|A_2|/|A_0|$. 
Equation (\ref{epsp2}) can be rewritten
\begin{equation}
\frac{\epsilon^\prime}{\epsilon}
= - \frac{i\omega e^{i(\delta_2 - \delta_0-\Phi_{\epsilon})}}
{\sqrt{2}|\epsilon|{\rm Re}\,A_0}
\left\{
{\rm Im}\, A_0^{(0)} (1 - \tilde\Omega_{\eta + \eta'})
- \frac{1}{\omega} {\rm Im}\, A_2^{(0)}\right\}
\,
\end{equation}
where 
\begin{equation}
\tilde\Omega_{\eta + \eta'}
= 
- \frac{\sqrt{2}}{3\omega}
\frac{|A_{\rm IB}^{00}|}{|A_0|}
\;.
\end{equation}
This is identical to Eq.~(\ref{epsdefold}) as $\tilde\Omega_{\eta + \eta'}=
\Omega_{\eta + \eta'}$.
In \opfour this simple interpretation of isospin-violating
contributions in $\Omega_{\rm IB}$
as modifications of $\omega$ does not carry as
$\xi_{+-}\ne \xi_{00} \ne \xi_0$ in general. The
interpretation also fails if $(27_L,1_R)$ operators
are included in the description of isospin-violating
effects.

\section{Conclusions}

We have established a framework for the
analysis of $K\ra\pi\pi$ decays in the presence of strong-interaction
isospin violation, so that the ``true''
$|\Delta I|=1/2$ and $|\Delta I|=3/2$ amplitudes can be assessed.
In particular, using unitarity arguments, 
we have shown that Watson's theorem, namely, 
the 
parametrization of Eq.~(\ref{newparam}), is 
appropriate to ${\cal O}((m_d - m_u)^2)$
to all orders of chiral perturbation theory. 
If we accept, as per Ref.~\cite{cirigliano},
that electromagnetic effects do not alter the structure of
Eq.~(\ref{newparam}), 
we can enlarge our analysis of $K\ra\pi\pi$ decays
in ${\cal O}(m_d - m_u)$ to include electromagnetic
effects as well. 
Incorporating the electromagnetic corrections of 
Ref.~\cite{cirigliano} and the $\delta_0-\delta_2$ phase
shift of Ref.~\cite{meissner}, we are unable to fit the
$K\ra\pi\pi$ branching ratio data with 
effective $(8_L,1_R)$ and $(27_L,1_R)$
low-energy constants in the framework of
chiral perturbation theory, as our fits require 
the existence of intolerably large, higher-order
corrections. Our failure, in retrospect, is predicated
by the observation that the empirical value of the 
$|\Delta I|=5/2$ amplitude, determined by the value of the 
$\delta_0-\delta_2$ phase shift,
is much larger and of opposite sign to the 
electromagnetically generated 
$|\Delta I|=5/2$ amplitude computed by Ref.~\cite{cirigliano}
in either chiral perturbation theory or in 
their dispersive matching approach.
Although our results suggest that our phenomenological
analysis is incomplete, that is, that missing electromagnetic effects 
likely exist, it is clear that the value of 
$A_2/A_0$ --- the ``true'' ratio of the $|\Delta I|=3/2$
to $|\Delta I|=1/2$ amplitudes --- is quite 
uncertain, as it is sensitive to the
inclusion of isospin-violating effects. 

Turning to an analysis of \epsrat in the presence of 
isospin violation, and 
applying the parametrization of Eq.~(\ref{newparam}), we 
find that an empirical $|\Delta I|=5/2$ amplitude of the magnitude
we have found generates a significant decrease in 
the Standard Model prediction 
of \epsrat --- although this decrease has a considerable
uncertainty, quantified through the errors in the
$K\ra\pi\pi$ branching ratios and the $\delta_0-\delta_2$ 
phase shift. 

\noindent {\bf Acknowledgments} 
We are grateful to J.~F. Donoghue and H.~R. Quinn
for helpful comments and discussions. 
The work of S.G. and G.V. was supported in
part by the DOE under contract numbers 
DE-FG02-96ER40989 and DE-FG02-92ER40730, respectively.
We are grateful to the Center for the Subatomic Structure of Matter at the
University of Adelaide, Brookhaven Theory Group, the Fermilab Theory Group, 
and the SLAC Theory Group for hospitality during the completion of this work. 

\section{Appendix}

We wish to consider how \opfour effects impact
the parametrization of Eq.~(\ref{paramotwo}). 
We
find by explicit calculation that 
the \opfour contributions of the weak, chiral Lagrangian of
Ref.~\cite{kambor} can be reorganized into
\begin{eqnarray}
&&A_{K^0\ra\pi^+\pi^-}=
\sqrt{2}C i ( 1 + \frac{2}{\sqrt{3}} \epsilon_1)
\left(\tilde g_8 + \tilde g_{27}^{(1/2)} + \tilde g_{27}^{(3/2)} + \frac{1}{2} 
\delta \tilde A_{5/2}^{\rm h.o.}\right) \non\\
&&A_{K^0\ra\pi^0\pi^0}=
\sqrt{2}C i ( 1 + \frac{2}{\sqrt{3}} \epsilon_1)
\left(\tilde g_8 + \tilde g_{27}^{(1/2)} - 2 \tilde g_{27}^{(3/2)} 
- \frac{2\epsilon_2}{\sqrt{3}}(\tilde g_8 + 
6 \tilde g_{27}^{(1/2)} - 3\tilde g_{27}^{(3/2)})
- 
\delta \tilde A_{5/2}^{\rm h.o.}\right) 
\label{paramofour}\\
&&A_{K^+\ra\pi^+\pi^0}=
C i ( 1 - \frac{2}{\sqrt{3}} \epsilon_1)
 \left(3\tilde g_{27}^{(3/2)}
+ \frac{\epsilon_2}{\sqrt{3}}(2 \tilde g_8 + 12 \tilde g_{27}^{(1/2)} + 
3 \tilde g_{27}^{(3/2)})
- \delta \tilde A_{5/2}^{\rm h.o.}\right) \;,
\non
\end{eqnarray}
where the effects of the higher-order weak counterterms
are lumped into the effective constants 
$\tilde g_8$, $\tilde g_{27}^{(1/2)}$, $\tilde g_{27}^{(3/2)}$, and a new 
$|\Delta I|=5/2$ contribution 
$\delta \tilde A_{5/2}^{\rm h.o.}$, which is of order 
$D_i B_0(m_d - m_u)$, where $D_i$ is a \opfour counterterm of 
$(27_L,1_R)$ character.  
Were 
$\delta \tilde A_{5/2}^{\rm h.o.}=0$ and $\epsilon_1=\epsilon_2=\epsilon_8$,
we would recover the parametrization 
of Eq.~(\ref{paramotwo}).
In Eq.~(\ref{paramofour}), we have
explicitly separated the strong-interaction isospin violation 
which emerges from meson mass differences, namely 
$m_{K^0,K^+}^2$, from that generated by $\pi^0-\eta$ mixing. The parameters 
$\epsilon_1$ and $\epsilon_2$ 
denote these two respective sources of isospin violation. 
Note that isospin-violating effects beyond $\pi^0$-$\eta$ mixing,
as discussed in Ref.~\cite{sggv}, are embedded
in $\tilde g_{27}^{(3/2)}$ and $\tilde g_{27}^{(1/2)}$.
In ${\cal O}(p^2)$, 
$\epsilon_1$ and $\epsilon_2$ 
are given by $\sqrt{3}(m_d - m_u)/(4(m_s - \hat{m}))$. 
In ${\cal O}(p^4)$, $\epsilon_2$ is modified by 
$\pi^0-\eta^\prime$ mixing, as realized by
the coefficients of the \opfour strong chiral 
Lagrangian~\cite{ecker}. 
Note that the cancellation of the $\epsilon_8 g_8$
contribution to the $K\ra \pi^0\pi^0$ amplitude found in \optwo 
no longer occurs if $\epsilon_1 \ne \epsilon_2$. 
Working consistently to ${\cal O}(m_d - m_u)$, and including
electromagnetic effects, we
find that Eq.~(\ref{paramofour}) implies
\begin{eqnarray}
&&x=\frac{\sqrt{2} r^{(3/2)} }{1 + r^{(1/2)}}\left(1 - 
\frac{2}{3\sqrt{3}}
\frac{(3\epsilon_1  - \epsilon_2  
+ 3r^{(3/2)} \epsilon_2  
+ 3r^{(1/2)} (\epsilon_1 - 2 \epsilon_2)
)}{1+ r^{(1/2)}}
- 
\frac{h_1 C_{em} (2 C_{+-} + C_{00})}{3(1+ r^{(1/2)})}
 \right) 
\label{loxemho} \\ 
&& \qquad + \frac{1}{15}\sqrt{\frac{2}{3}}
\frac{(10\epsilon_2 - r^{(3/2)}(6\epsilon_1 + 3\epsilon_2)   
+ 60 r^{(1/2)}\epsilon_2 )}{1 + r^{(1/2)}}
+ \frac{\sqrt{2}}{5}
\frac{h_1 C_{em} (2(C_{+-} - C_{00}) + 3C_{+0})}{3(1+ r^{(1/2)})}
\non
\end{eqnarray}
and
\begin{equation}
y=\frac{\sqrt{2}}{5}\left( \frac{\sqrt{3} r^{(3/2)}
(4 \epsilon_1 - 3\epsilon_2) 
+ h_1 C_{em}(C_{+-} - C_{00} - C_{+0})}
{1 + r^{(1/2)}}\right) 
+ \frac{1}{\sqrt{2}} 
\frac{(\delta \tilde A_{5/2}^{\rm h.o.}/\tilde g_8)}{1 + r^{(1/2)}}
\;. 
\label{loyemho}
\end{equation}
We have defined 
$r^{(3/2)}\equiv \tilde g_{27}^{(3/2)}/\tilde g_8$ 
and $r^{(1/2)}\equiv \tilde g_{27}^{(1/2)}/\tilde g_8$, and
the parameter $h_1\equiv g_8/\tilde g_8$. We estimate 
\begin{eqnarray}
\frac{\delta \tilde A_{5/2}^{\rm h.o.}}{\tilde g_8}
&\sim& 
\left(
\frac{\tilde g_{27}^{(3/2)}}{\tilde g_8}
\right)
\left(
\frac{g_{27}^{(3/2)}}{\tilde g_{27}^{(3/2)}}
\right)
\left(
\frac{4\epsilon_2}{\sqrt{3}}
\right)
\left(
\frac{B_0(m_s - \hat{m})}{\Lambda_{\chi SB}^2} \right)
\non \\
&\equiv& (0.52) h_2 r^{(3/2)} \epsilon_2 \;,
\end{eqnarray}
where $B_0(m_s - \hat{m})/\Lambda_{\chi SB}^2\sim 0.23$. 
We expect the parameters $h_1$ and $h_2$ to be of order unity. 
Higher-order effects in the weak chiral Lagrangian
serve to make $\tilde g_{27}^{(1/2)}\ne \tilde g_{27}^{(3/2)}/5$
--- the term $D_6$, e.g., in the \opfour weak, chiral Lagrangian
of Ref.~\cite{kambor} generates such an inequality.
Consequently, we expect from dimensional analysis
\begin{equation}
\frac{\delta \tilde g_{27}^{1/2}}{\tilde g_{27}^{3/2}}\equiv
\frac{\tilde g_{27}^{(1/2)} - \tilde g_{27}^{(3/2)}/5}{\tilde g_{27}^{(3/2)}}
\sim 
\left(
\frac{g_{27}^{(3/2)}}{\tilde g_{27}^{(3/2)}}
\right)\left(
\frac{ B_0(m_s - \hat{m})}{\Lambda_{\chi SB}^2} \right)
\equiv
0.23 h_3
\;, 
\label{brkdim}
\end{equation}
where the parameter $h_3$ ought be of order unity. A model estimate of 
$\delta \tilde g_{27}^{1/2}/\tilde g_{27}^{3/2}$
suggests that it is less than $30\%$~\cite{pich86}.
Isospin-violating contributions, ignored in Eq.~(\ref{brkdim}),
also contribute to $\delta \tilde g_{27}^{1/2}/\tilde g_{27}^{3/2}$;
the largest terms are typified by $B_0 (m_d - m_u) E_i$, where $E_i$ is 
an \opfour counterterm of $(8_L,1_R)$ in character, and thus generate,
crudely, an additional $\sim 10\%$ effect. The value of 
$(r^{(1/2)} -  r^{(3/2)}/5)/r^{(3/2)}$ found in Table~\ref{tableall}
far exceeds the estimate of Eq.~(\ref{brkdim}). 
We thus wish to see whether plausible choices of $h_1$, 
$h_2$, and $\epsilon_2$ can serve to reduce
the SU(3)$_f$ breaking of the relation 
$r^{(3/2)}=r^{(1/2)}/5$ found in Table~\ref{tableall} 
to a plausible level. 

We explore how the values of 
$r^{(3/2)}$ and $r^{(1/2)}$ vary as a function of 
$\epsilon_2$, $h_1$, and $h_2$ in Table~\ref{tablefix}.
We fix $\epsilon_1=0.0106 \pm 0.0008\;$~\protect{\cite{leut96}}
and choose $\delta_0-\delta_2 =51^\circ$. The latter is determined 
by the central value of $45^\circ$ given in Ref.~\cite{meissner}
plus $6^\circ$, the +1$\sigma$ excursion permitted.
We estimate that $h_1$ could be as small as $0.5$, and
we choose two different values for $\epsilon_2$: 
we use the result determined from the \opfour strong
chiral Lagrangian of Ref.~\cite{ecker} as well as the
estimate $\epsilon_2 = 2 \epsilon_1 \pm \epsilon_1$. 
The central value and its error assigned
to $\epsilon_2$ in this latter estimate is rather
generous; we observe that electromagnetic effects,
not included in Ref.~\cite{cirigliano}, can enhance
the $\pi^0$-$\eta,\eta^\prime$ mixing angle slightly~\cite{dght}. 
Despite our efforts, a value of $h_2 \sim -25$ or larger
is required to make the SU(3)$_f$ breaking of 
$(r^{(1/2)} -  r^{(3/2)}/5)/r^{(3/2)}$ no more than 100\%. 
Interestingly, replacing the estimates of the
electromagnetic corrections in the dispersive matching
approach with those determined in chiral perturbative theory
does increase the errors in the determined values of 
$r^{(3/2)}$ and $r^{(1/2)}$, but not sufficiently
to reduce the value of $h_2$ substantially. 
It seems unlikely that strong-interaction 
isospin-violating effects can resolve the difference
between the empirical value of $y$ predicated by 
a phase shift $\delta_0 - \delta_2 \sim 45^\circ$
and the electromagnetic effects computed in Ref.~\cite{cirigliano}.  

\begin{table}[htb]
\begin{center}
\caption{
The values of $r^{(1/2)}$ and $r^{(3/2)}$ 
determined by fitting Eqs.(\protect{\ref{loxemho}})
and (\protect{\ref{loyemho}}) to 
the empirically determined $x$ and $y$, 
resulting from the phase shift difference, $\delta_0-\delta_2$. 
Solutions yielding $A_2/A_0 \gtrsim 1$
have been omitted. 
The parameter $\epsilon_1=0.0106 \pm 0.0008\;$~\protect{\cite{leut96}}
throughout. No errors are assigned to the $h_1$ and 
$h_2$ parameters. 
Note that (C) and (D) denote the electromagnetic 
corrections of Ref.~\cite{cirigliano}
as computed in chiral perturbation theory (C) and in 
the ``dispersive
matching'' (D) approach. The ratio 
$A_2/A_0\equiv \sqrt{2}r^{(3/2)}/(1 + r^{(1/2)})$ does
include $m_d\ne m_u$ effects through
the absorbed \opfour counterterms.
\smallskip
}
\begin{tabular}{cccc}
\hline
$\delta_0-\delta_2  = 51^\circ \pm 6^\circ $ &  
$\epsilon_2=2\epsilon_1\pm \epsilon_1$   & 
$h_1=1 \quad h_2=-1$ \\ 
\hline
$r^{(3/2)}$  & $r^{(1/2)}$ & $A_2/A_0$ & $\hbox{em}$ \\
 $   0.044 \pm   0.026 $ &
 $  -1.49 \pm   0.77 $ &
 $  -0.13 \pm   0.19 $ & $\hbox{C}$ \\ 
 $   0.047 \pm   0.027$  &
 $  -1.59 \pm   0.72 $ &
 $  -0.11 \pm   0.14 $ & $\hbox{D}$ \\ 
\hline
$\delta_0-\delta_2  = 51^\circ \pm 6^\circ $ &  
$\epsilon_2=2\epsilon_1\pm \epsilon_1$   & 
$h_1=0.5 \quad h_2=-1$ \\ 
\hline
$r^{(3/2)}$  & $r^{(1/2)}$ & $A_2/A_0$ & $\hbox{em}$ \\
 $   0.035 \pm   0.016 $ &
 $  -1.19 \pm   0.35 $ &
 $  -0.26 \pm   0.44 $ & $\hbox{C}$ \\ 
 $   0.037 \pm   0.015 $ &
 $  -1.24 \pm   0.33 $ &
 $  -0.22 \pm   0.33 $ & $\hbox{D}$ \\ 
\hline
$\delta_0-\delta_2  = 51^\circ \pm 6^\circ $ &  
$\epsilon_2=2\epsilon_1\pm \epsilon_1$   & 
$h_1=0.5 \quad h_2=-25$ \\ 
\hline 
$r^{(3/2)}$  & $r^{(1/2)}$ & $A_2/A_0$ & $\hbox{em}$ \\
 $   0.0296 \pm   0.0092 $ &
 $  -0.36 \pm   0.48 $ &
 $   0.065 \pm   0.069 $ & $\hbox{C}$ \\ 
 $   0.0303 \pm   0.0086 $ &
 $  -0.39 \pm   0.44 $ &
 $   0.070 \pm   0.070 $ & $\hbox{D}$ \\ 
\hline
$\delta_0-\delta_2  = 51^\circ \pm 6^\circ $ &  
$\epsilon_2= 0.0133\pm  0.0025$~\protect{\cite{ecker}}
 & 
$h_1=0.5 \quad h_2=-25$ \\ 
\hline 
$r^{(3/2)}$  & $r^{(1/2)}$ & $A_2/A_0$ & $\hbox{em}$ \\
 $   0.0289 \pm   0.0064 $ &
 $  -0.80 \pm   0.27 $ &
 $   0.20 \pm   0.30 $ & $\hbox{C}$ \\ 
 $   0.016 \pm   0.011 $ &
 $  -1.046 \pm   0.090 $ &
 $  -0.48 \pm   0.62 $ & $\hbox{C}$ \\ 
 $   0.0300 \pm   0.0059 $ &
 $  -0.83 \pm   0.17 $ &
 $   0.25 \pm   0.28 $ & $\hbox{D}$ \\ 
 $   0.0174 \pm   0.0059 $ &
 $  -1.063 \pm   0.072 $ &
 $  -0.39 \pm   0.33 $ & $\hbox{D}$ \\ 
\hline
$\delta_0-\delta_2  = 51^\circ \pm 6^\circ $ &  
$\epsilon_2=2\epsilon_1\pm \epsilon_1$   & 
$h_1=0.5 \quad h_2=-50$ \\ 
\hline 
$r^{(3/2)}$  & $r^{(1/2)}$ & $A_2/A_0$ & $\hbox{em}$ \\
 $   0.0215 \pm   0.0097 $ &
 $   0.02 \pm   0.57 $ &
 $   0.030 \pm   0.029 $ & $\hbox{C}$ \\ 
 $   0.0220 \pm   0.0093$ &
 $   0.00 \pm   0.54 $ &
 $   0.031 \pm   0.029 $ & $\hbox{D}$ \\ 
\hline
$\delta_0-\delta_2  = 51^\circ \pm 6^\circ $ &  
$\epsilon_2= 0.0133\pm  0.0025$~\protect{\cite{ecker}} &
$h_1=0.5 \quad h_2=-50$ \\ 
\hline 
$r^{(3/2)}$  & $r^{(1/2)}$ & $A_2/A_0$ & $\hbox{em}$ \\
 $   0.026 \pm   0.0013 $ &
 $  -0.37 \pm   0.72 $ &
 $   0.058 \pm   0.067 $ & $\hbox{C}$ \\ 
 $   0.0260 \pm   0.0016 $ &
 $  -0.41 \pm   0.63 $ &
 $   0.063 \pm   0.069 $ & $\hbox{D}$ \\ 
\hline
\end{tabular}
\label{tablefix}
\end{center}
\end{table}


\begin{thebibliography}{999}
\bibitem{foot} Note that the properly symmetrized state 
$| \pi^-\pi^+ \rangle_{\rm sym} \equiv 
(| \pi_1^+\pi_2^- \rangle +  | \pi_1^-\pi_2^+ \rangle)/\sqrt{2}
= \sqrt{2} |\pi^+\pi^-\rangle$ ought appear throughout. The extra
factor of $\sqrt{2}$ in the definition of 
$|\pi^-\pi^+ \rangle_{\rm sym}$ implies 
that the branching ratio into $K\ra\pi^+\pi^-$ is 
a factor of 2 larger than suggested by Eq.~(\ref{isodecomp}). 


\bb{oddpar} The parity of a three-$\pi$ state, with 
coordinates $\vec{r}_1, \vec{r}_2,$ and
$\vec{r}_3$, is determined by the function
$-(\vec{r}_2 - \vec{r}_1)^{l_1}
(\vec{r}_3 - (\vec{r}_1 + \vec{r}_2)/2)^{l_2}$. For the
$J=0$ state, $l_1=l_2=l$, so that the $J=0$ state is
of odd parity.
\bibitem{roy} The only inelastic channel in the vicinity
of $\sqrt{s} \sim m_K$ is the transition 
of $2\pi \ra 4\pi$ with a threshold of 
$\sqrt{s} \simeq 0.56$ GeV; this reaction
is also strongly phase-space suppressed near threshold, 
so that no significant 
inelasticity exists for $\sqrt{s} < 0.8$ GeV~\cite{ananth}.
\bibitem{neglectem2} This procedure is consistent with the 
parametrization of Eq.(3) in the first paper of Ref.~\cite{cirigliano}:
the electromagnetic effects may alter the values of $\delta_I$ or
of the amplitudes themselves, but the form of the parametrization
in Eq.~(\ref{newparam}) persists.
\end{thebibliography}
\end{document}